\documentclass[sigconf]{acmart}
\acmConference{}{}{}
\renewcommand\footnotetextcopyrightpermission[1]{} 
\makeatletter
\renewcommand\@formatdoi[1]{\ignorespaces}
\makeatother

\usepackage{algorithm2e}
\usepackage{amsmath}
\usepackage{caption}
\usepackage{subcaption}
\usepackage{appendix}
\usepackage{listings}
\usepackage{fancyvrb}
\usepackage{color}
\usepackage[export]{adjustbox}

\newcommand{\norm}[1]{\left\Vert #1 \right\Vert}
\DeclareMathOperator{\dist}{dist}
\newcommand\sbullet[1][.5]{\mathbin{\vcenter{\hbox{\scalebox{#1}{$\bullet$}}}}}

\AtBeginDocument{%
  \providecommand\BibTeX{{%
    \normalfont B\kern-0.5em{\scshape i\kern-0.25em b}\kern-0.8em\TeX}}}

\begin{document}

\title{A SVBRDF Modeling Pipeline using Pixel Clustering}

\author{Bo Li}
\email{li\_bo@pku.edu.cn}
\affiliation{%
  \institution{Peking University}
}

\author{Jie Feng}
\email{feng\_jie@pku.edu.cn}
\affiliation{%
  \institution{Peking University}
}

\author{Bingfeng Zhou}
\email{cczbf@pku.edu.cn}
\affiliation{%
  \institution{Peking University}
}


\renewcommand{\shortauthors}{Li et al.}

\begin{abstract}
We present a pipeline for modeling spatially varying BRDFs (svBRDFs) of planar materials which only requires a mobile phone for data acquisition. With a minimum of two photos under the ambient and point light source, our pipeline produces svBRDF parameters, a normal map and a tangent map for the material sample. The BRDF fitting is achieved via a pixel clustering strategy and an optimization based scheme. Our method is light-weight, easy-to-use and capable of producing high-quality BRDF textures.
\end{abstract}




\begin{teaserfigure}
 \centering
  \includegraphics[width=\textwidth]{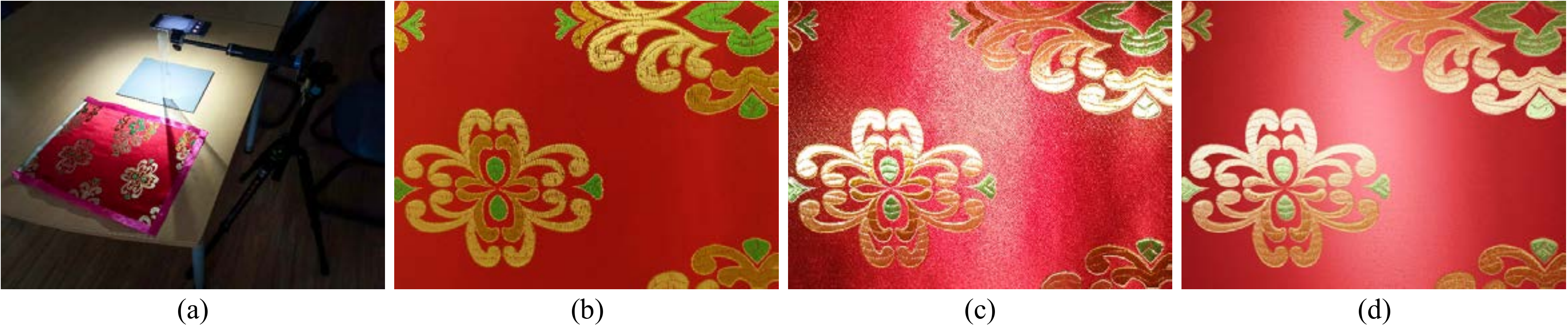}
  \vspace{-0.8cm}
  \caption{
    \label{fig:teaser}
      Using a simple consumer-level mobile camera setup (a), we take a photo of the target material under the ambient light (b) and one under the flash light (c). With a minimum of two input photos, our pipeline can generate high-resolution svBRDF textures, which can be faithfully rerendered under the original illumination (d).}
\end{teaserfigure}

\maketitle

\section{Introduction}
The need for real-world material modeling has grown rapidly in computer graphics. Although the theory of physically based rendering (PBR) has been thoroughly studied, its inverse procedure, recovering material appearance from rendered images or photos, remains an ill-posed problem. Moreover, real-world material usually has reflectance that changes with position, i.e., spatially varying BRDF (svBRDF) or bidirectional texture function (BTF)~\cite{DvGN*99}, that increases the difficulty for this inverse problem. In practice, parametric svBRDF models with textured parameters are often used. They are more compact in storage and flexible for rendering and editing. However, manual generation of svBRDF textures is a time-consuming work and requires parameter tuning. 

Recently, many light-weight solutions have been proposed, indicating capturing process with slight cost can result in fairly satisfactory svBRDFs~\cite{LKG*03,AWL15,ACGO18}. For simplicity, many similar researches focus on materials of simple planar geometry. These methods share the idea of redundancy of data, to be specific, the spatially varying material can be split into finite categories or represented by the combination of base materials. Limitations of the previous work include texture-like material assumption~\cite{AWL15}, inability for Fresnel effects~\cite{ACGO18}, etc. We proceed to light-weight svBRDF recovering solutions based on innovations of previous work.

In this paper, we introduce a novel svBRDF modeling pipeline for planar materials. Our proposed pipeline is efficient and easy-to-use, which requires only a minimum of two images for each material sample as input. No rigorous assumptions are made, such as the category of material or pattern repeating characteristics like procedural textures.

We adopt an intuitive BRDF model which is representative among physically based shading models. An iterative multi-stage optimization process for fitting model parameters is proposed, using simple loss functions that alleviates deliberate design of regulation terms. To reduce the complexity of the parameter optimization, a pixel clustering algorithm is introduced to effectively quantize material textures into finite clusters, according to both colors and local structures. This clustering algorithm is resistant to noise and preserves details of the material at the same time. Besides, for the purpose of asset creation in computer graphics, it is desirable if svBRDF models behave in accordance to real-world physical reflectance, decoupling the impact of image acquisition equipment, lighting conditions, etc. With a few additional calibration images under a fixed camera setting, our pipeline is capable of producing high-quality svBRDF textures without being affected by these irrelevant factors.

\section{Related Work}
Classical direct BRDF measurement methods rely on special-purpose equipments known as gonioreflectometers~\cite{War92, Foo97, DvGN*99, LFTW06} or other specifically designed devices~\cite{MSY07, DWT*02, AWL13}. In comparison, image-based material modeling approaches target to recover the reflectance and geometry of objects, and require only general-purpose cameras. Lensch et al.~\cite{LKG*03} use photos to recover geometry and reflectance by representing BRDF with a linear combination of basis BRDFs. Dong et al.~\cite{DWT10} proposes a manifold bootstrapping method, highlighting that material BRDF is a low-dimensional manifold formed by some representative BRDFs. AppGen ~\cite{DTPG11} is an interactive material modeling process from single image. It uses intrinsic image decomposition techniques to recover diffuse albedo and normal, then assigns specular properties guided by user supplied strokes and diffuse information. Aittala et al.~\cite{AWL15} take two photos under natural light and flash light, which inspires our pipeline input. They cut the image into tiles and utilize the similarity between different tiles, thus their solution is restricted to texture-like materials. They introduce elaborate regulation terms to guarantee smoothness and curl-free property of normals. Albert et al.~\cite{ACGO18} create svBRDF from mobile phone video, featuring video frame alignment and iterative subclustering strategy. 

With the increasing popularity of deep learning methods in computer vision and computer graphics, Convolutional Neural Networks (CNNs) have been implemented in appearance modeling questions. Aittala et al.~\cite{AAL16} use a neural style texture transfer strategy. A great difficulty in these supervised learning methods is the acquisition of labeled training data, i.e., photos with ground truth svBRDFs. Li et al.~\cite{LDPT17} train a CNN to approximate svBRDF map from a single image. To reduce need for manually labeled data during network training, they exploit a strategy called self-augmentation that render images with svBRDF maps predicted from unlabeled images to obtain new ground truth labeled data pairs. Deschaintre et al.~\cite{DAD*18} use procedural svBRDF to render ground truth images and augment training data with random perturbations. Li et al.~\cite{LSC18} split materials into several categories and introduces a classifier to assign weights for blending among different categories. Li et al.~\cite{LXR*18} use CNN to predict geometry (depth and normal), diffuse albedo and specular roughness with one image under uncontrolled conditions ("in-the-wild"). 

\section{Pipeline Overview}
\begin{figure}[htb]
      \centering
      \includegraphics[width=1.0\linewidth]{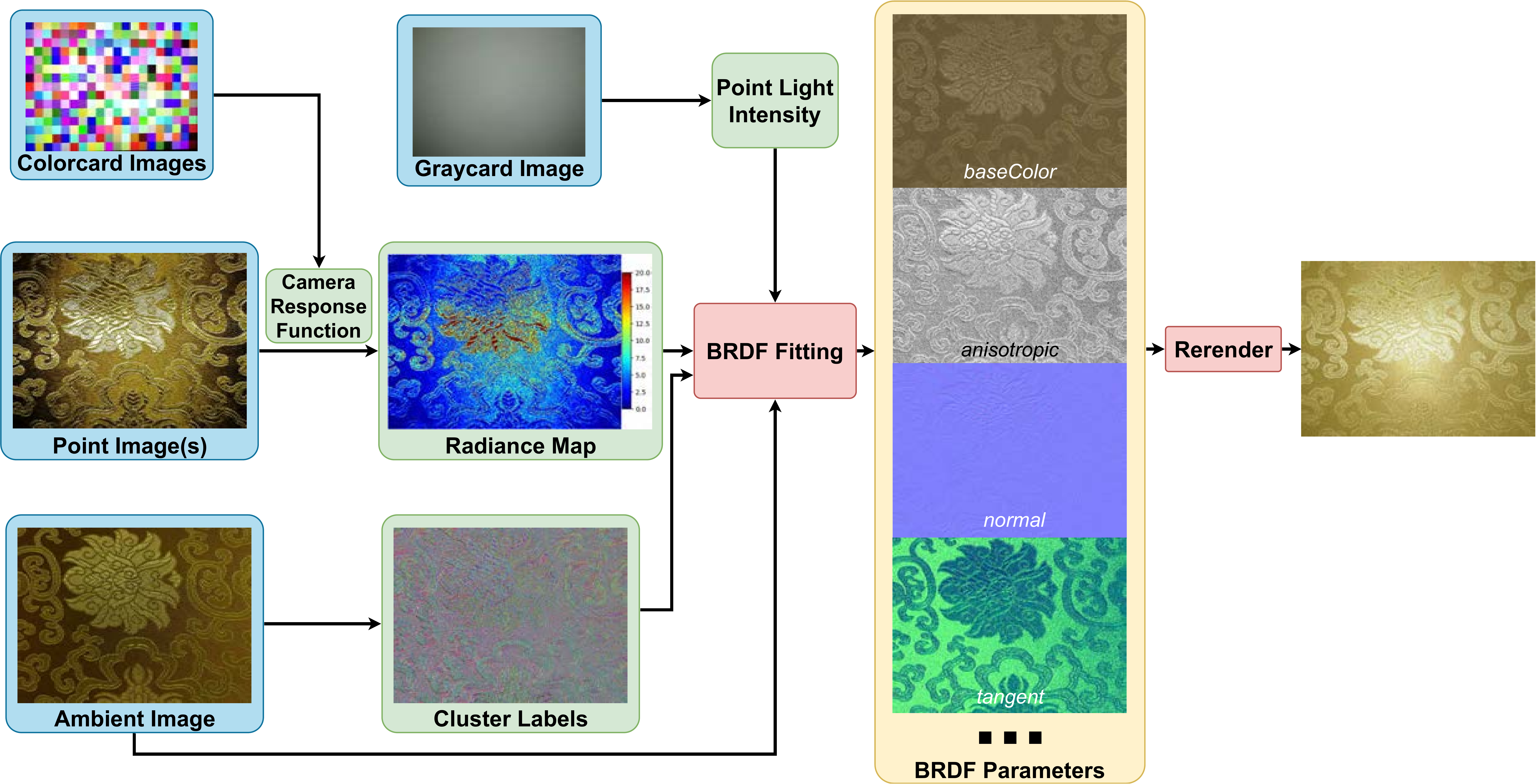}
      %
      \caption{\label{dataflow}
            Data flow of our pipeline. Blue background images are inputs of the pipeline. Nearly ten color card images are used, and multiple point images can be involved in recovering the radiance map. Green intermediate quantities are produced to fit svBRDF results with yellow background.}
\end{figure}

In this paper, we present an appearance modeling pipeline with a small number of photos (Figure~\ref{dataflow}). For each material sample, we take two photos sharing a fixed position: one under natural, ambient lighting (referred as \textit{ambient image}) and the other illuminated by a point light (\textit{point image}). The ambient image is utilized to discover similar parts on the material and classify the pixels into finite clusters (Section~\ref{clustering}). It also helps to extract rich details of local contrast induced by bumps, forming a height (normal) map (Section~\ref{heightmap}). The point image is set as target for optimizing BRDF parameters (Section~\ref{global},~\ref{sv}). Due to the limited dynamic range of digital cameras, point images of materials with strong specular highlight may be taken under different exposure times, akin to high-dynamic-range imaging (HDRI).

Some extra images, known as \textit{calibration images}, may be fed to the pipeline for eliminating factors that interfere with results. They include a series of images of a color chart under different exposures, which are employed to recover the camera response function that maps pixel value to real-world radiance. Besides, an image of a gray card commonly used in photography provides a reference to calculate the intensity of point light in the scene. 

Hence, with the radiance map derived from the point image, the clustering information from the ambient image, and the calibrated light intensity and camera response curve, BRDF fitting is solved as an optimization problem. The results of our pipeline are high-resolution svBRDF maps, bump maps and two global BRDF parameters. These maps and parameters can be applied to render the planar material under novel lighting and viewing, or mapped to 3D models for augmenting their appearance in PBR applications.

In the following subsections, we first introduce the hardware setup of our pipeline and the rendering model. Then the details of preprocessing and svBRDF modeling is given in Section~\ref{cali-pre} and Section~\ref{fitting-brdf}. Finally we present our experiment results and analysis in Section~\ref{experiment-analysis}.


\subsection{Photographic Hardware and Coordinate System}
\begin{figure}[htb]
            \centering
            \includegraphics[width=0.8\linewidth]{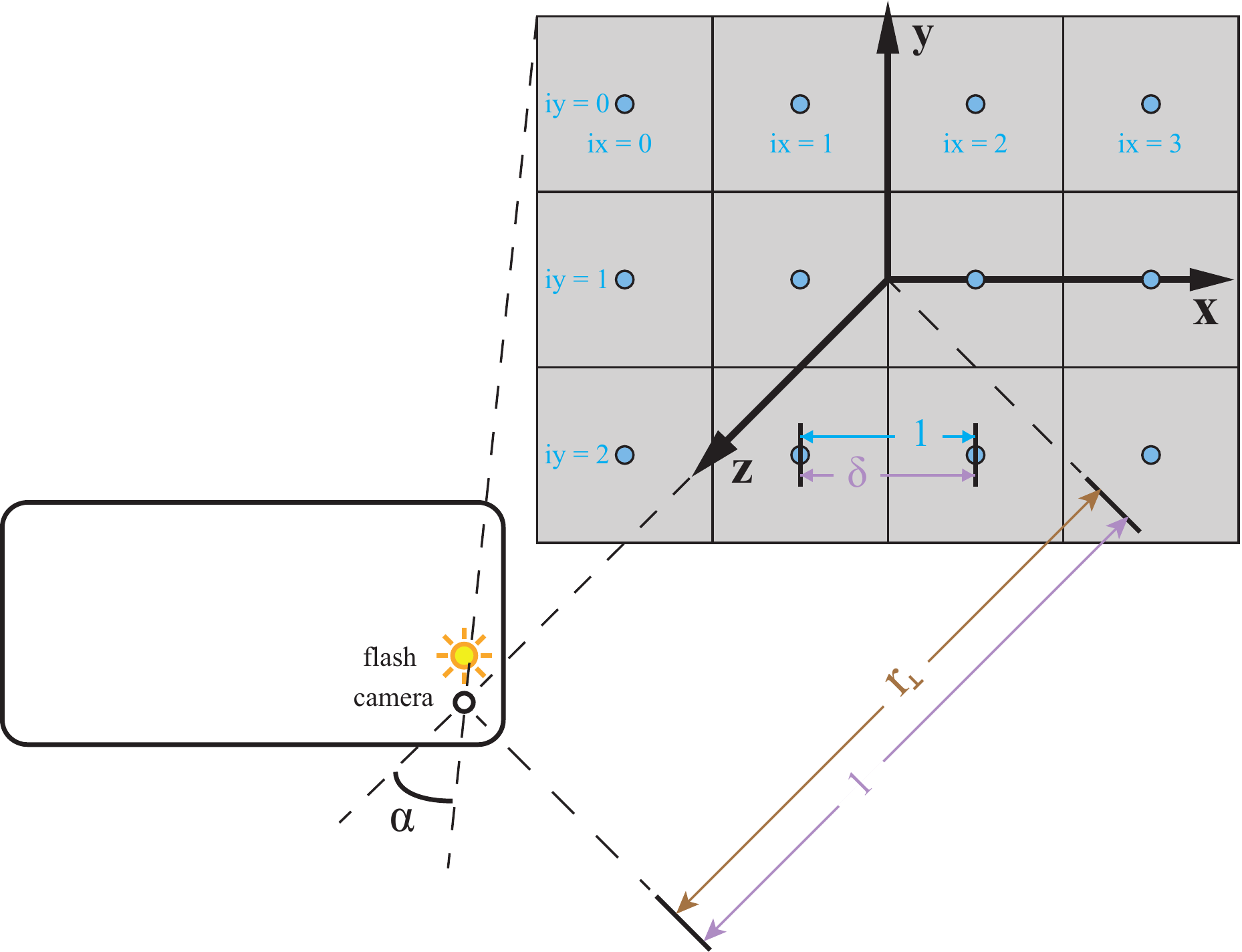}
      \caption{\label{coord}
      Illustration of the coordinate system. Distances are marked with three colors to distinguish measurement in different spaces. Cyan: image pixel indices; purple: our assumed coordinate space; brown: real world.}
\end{figure}

In our method, we require only a consumer-level camera with a light source that is small enough to be considered as a point light. Similar to~\cite{AWL15}, we use a smartphone with flash light as our image capturing device. With this hardware setup, we conduct reflectance fitting and rendering in a right-handed normalized coordinate system (Figure~\ref{coord}). For convenience, the material is assumed to lie in XY plane. The camera is at $\mathbf{p_o} = [0, 0, 1]$ on Z-axis, and its projection on the material plane is just the origin of the system and the center of photos at the same time. Note that distance from camera to XY plane in real world is $r_{\perp}$, which is recorded by hand and used in calibration (Section~\ref{graycard}). Point light position $\mathbf{p_i}$ is slightly apart from $\mathbf{p_o}$ because of displacement from mobile phone camera to flash light. In our implementation, $\mathbf{p_i}$'s pixel index is the weighted position of 10\% pixels with greatest grayscale values. Finally, to locate a pixel in the assumed coordinate system, we need to know the proportion between image indexing space and our coordinate space. To be exact, let one pixel's offset corresponds to $\delta$ unit displacement in X/Y direction (shown in Figure~\ref{coord}). To determine $\delta$, we utilize 35 mm equivalent focal length ($f_{35}$) in photo EXIF metadata. $f_{35}$ decides angle of view (AOV) and $\delta$ can be inferred accordingly, such that hard-coding AOV for specific camera can be avoided. The diagonal half AOV $\alpha$ is computed by the following formula~\cite{CIPA12, CIPA18}:
\begin{equation}
      \alpha = \arctan{\frac{d_{35}}{2f_{35}}} 
\end{equation}
Where $d_{35}$ is the diagonal size of 135 format film, about 43.3 mm. Considering that perpendicular distance from camera to material plane is 1, $\delta$ is then calculated by
\begin{equation}
      \delta = \frac{2\tan\alpha}{\sqrt{s_x^2 + s_y^2}} = \frac{d_{35}}{f_{35}\sqrt{s_x^2 + s_y^2}} 
\end{equation}
where $s_x$, $s_y$ correspond to image size.

\subsection{Rendering Model}
\label{render}
    Illuminated by a point light, the rendering equation for our system is in a simple point-wise form:
    
    \begin{equation}
            \label{render-model}
            L(\mathbf{p}) = f(\mathbf{x}(\mathbf{p}), \mathbf{\omega}_i, \mathbf{\omega}_o) \frac{E}{r(\mathbf{p})^2} \cos\theta_i
    \end{equation}
    where $\mathbf{p}$ indicates pixel position, $\omega_i$ and $\omega_o$ are unit vectors pointing to $\mathbf{p_i}$ and $\mathbf{p_o}$, $f$ is BRDF. Note that although scene radiance $L$ is proportional to irradiance arriving at a pixel of camera sensor, this proportionality factor often varies across different pixels, which partially leads to vignetting \cite{KMH95}. For simplicity, we just treat pixel irradiance as scene radiance, all noted as $L$. $E$ is point light's intensity, $r(\mathbf{p}) = \norm{\mathbf{p_o} - \mathbf{p}}$ is distance to point light, and $\cos\theta_i = \mathbf{n}(\mathbf{p}) \cdot \mathbf{\omega}_i$ is cosine of incident angle. Given two images under ambient and point light, our goal is to recover a set of BRDF parameters $\mathbf{x}$, normals $\mathbf{n}$ and tangents $\mathbf{t}$ for each pixel. 
    
    We apply a simplified version of Disney "principled" BRDF model, which is controlled by intuitive, comprehensible parameters bounded between 0-1 \cite{Bur12}. Many variants of this model are widely adopted in industry~\cite{Kar13, Lag14}. We simplified the original implementation, and choose the following parameters:
    \begin{itemize}
      \item \textit{baseColor} surface color related to albedo, affects both diffuse and specular lobe. 
      \item \textit{metallic} a blend between dielectric and metallic model.
      \item \textit{specular} controls the strength of specular reflection.
      \item \textit{specularTint} tints specular color from white to \textit{baseColor}. 
      \item \textit{roughness} affects zenithal angular response for both diffuse and specular lobe.
      \item \textit{anisotropic} extent of anisotropy. 0 means the model just degenerates to an isotropic one.
      \end{itemize}  
   A complete description can be found in the appendix. The ambient image is set as initial default value for \textit{baseColor}, 0.5 for \textit{specular} and \textit{roughness}, and 0 for the rest.
    
\section{Calibration and Preprocessing}
\label{cali-pre}
There are two problems hindering acquisition real reflectance of materials: 1. \emph{nonlinearity} of pixel values (i.e., doubling captured radiance does not result in doubling pixel values stored in photo); 2. \emph{reciprocity} between lighting and reflectance (i.e., they can be multiplied and divided by a same factor without affecting final render result). To deal with the two difficulties, we exploit the following calibration methods.
\subsection{Camera Response Curve Recovery}
\begin{figure*}[tbp]
      \centering
      \begin{subfigure}{0.19\linewidth}
            \centering
            \includegraphics[width=\linewidth]{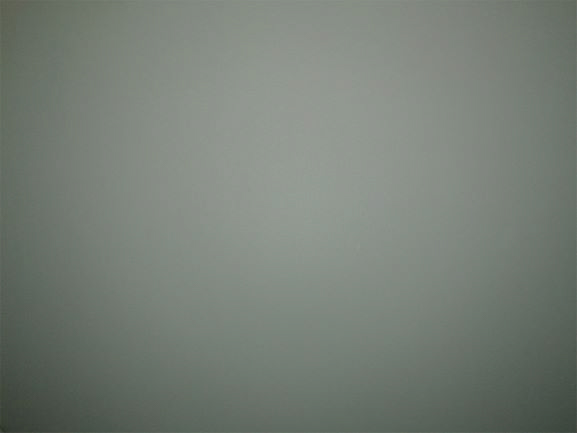}
            \caption{}
      \end{subfigure}
      \hfill
      \begin{subfigure}{0.19\linewidth}
            \centering
            \includegraphics[width=\linewidth]{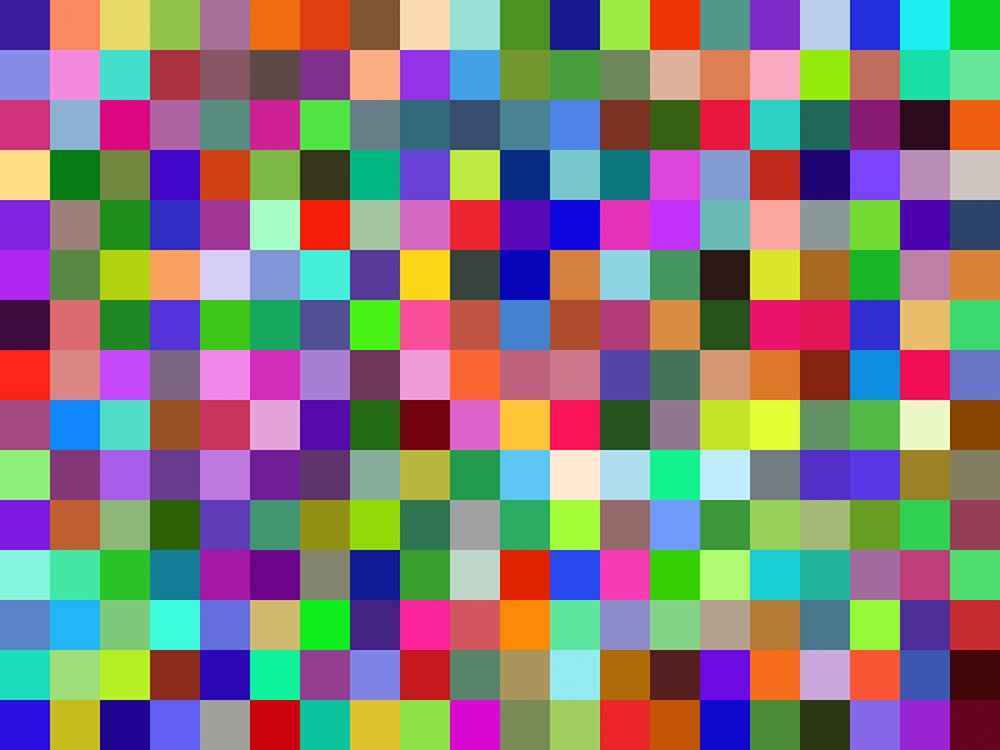}
            \caption{}
      \end{subfigure}
      \hfill
      \begin{subfigure}{0.19\linewidth}
            \centering
            \includegraphics[width=\linewidth]{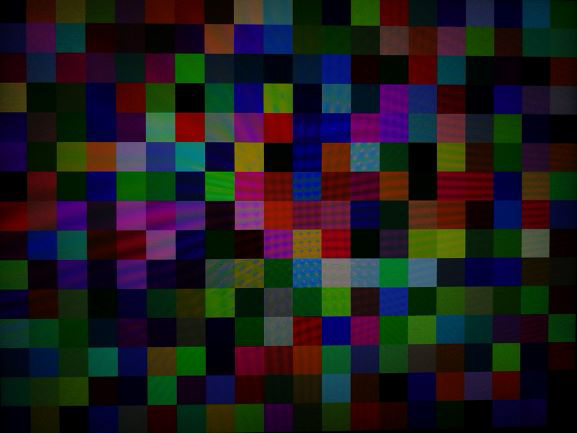}
            \caption{}
      \end{subfigure}
      \hfill
      \begin{subfigure}{0.19\linewidth}
            \centering
            \includegraphics[width=\linewidth]{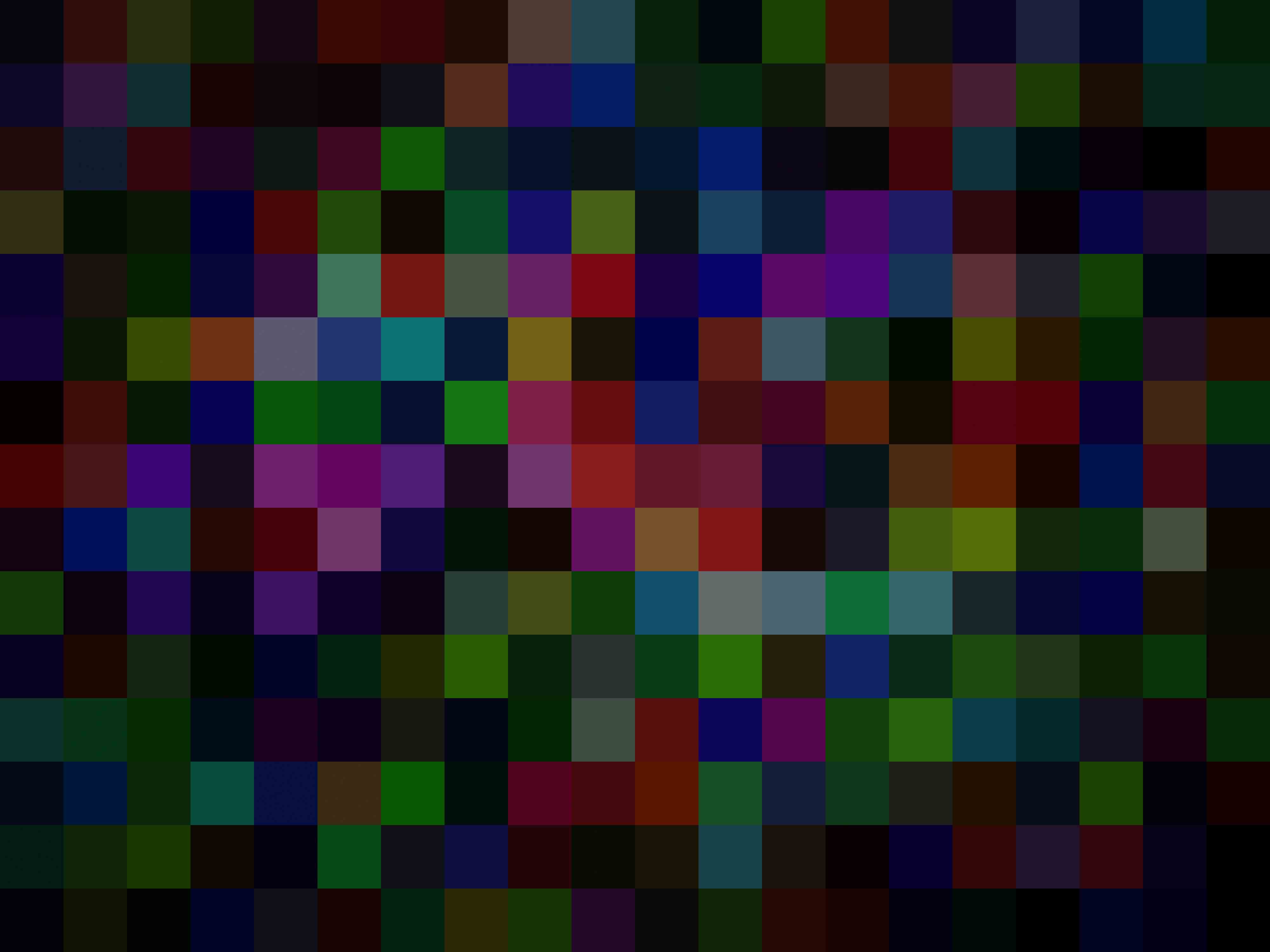}
            \caption{}
      \end{subfigure}
      \hfill
      \begin{subfigure}{0.22\linewidth}
            \centering
            \includegraphics[width=\linewidth, height=0.75\linewidth]{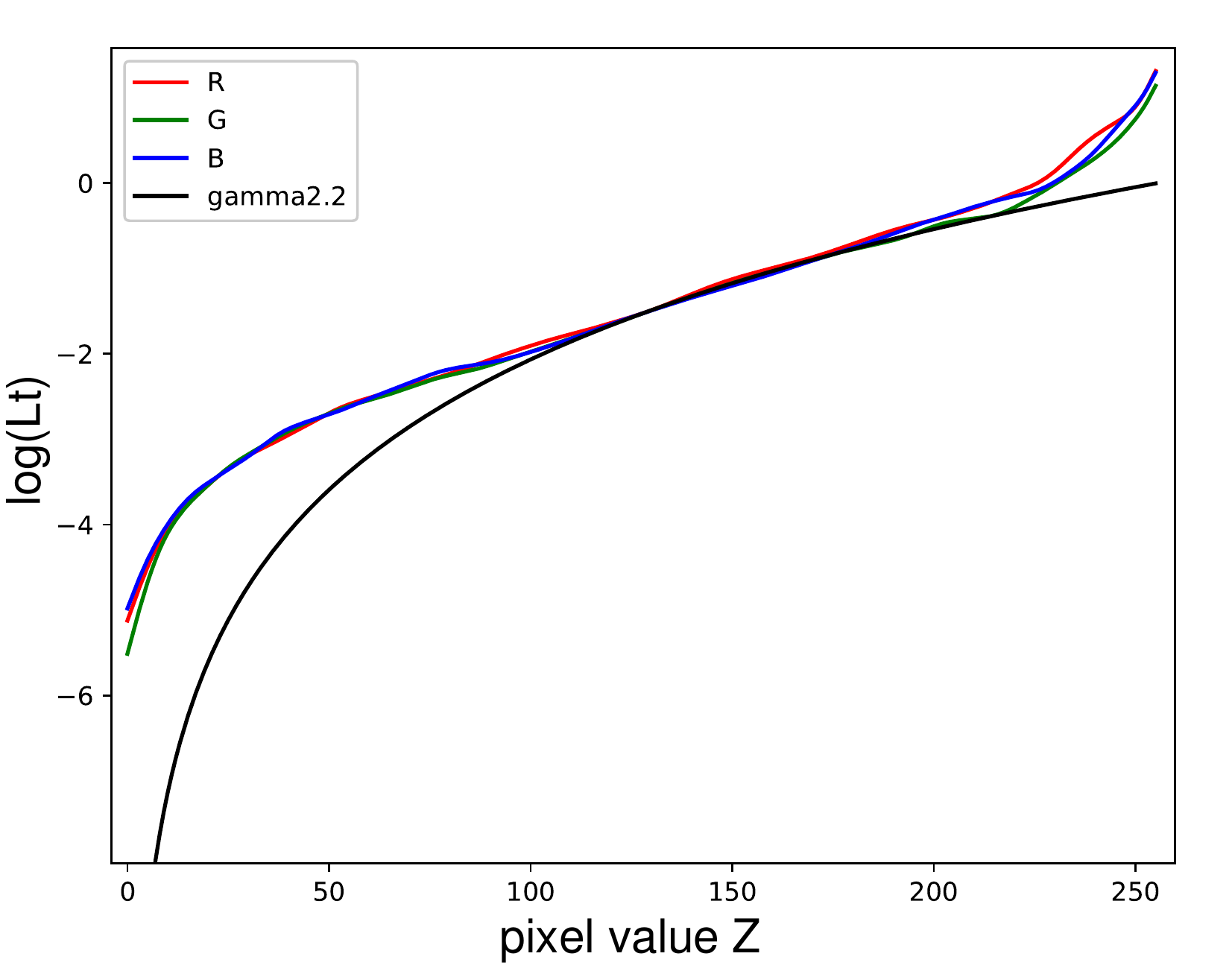}
            \caption{}
      \end{subfigure}
      \caption{\label{calibration}
            Images used in calibration. (a) A photo of the gray card. (b) A generated color card. (c) We display the virtual color card on a screen and capture images occupied by colored tiles, with different exposure times. (d) Resize images in last step to 20$\times$15. This downsampling generates stable observations, forming data-fitting equations for response curve optimization in~\cite{DM97}. (e) Recovered response curve. } 
\end{figure*}

One common way of treating nonlinearity is to apply an inverse gamma correction (exponent 2.2) to compensate gamma encoding in sRGB color space~\cite{SACM96}. However, this treatment only removes nonlinearity introduced during image storing, but not digital filming system itself. Debevec et al.~\cite{DM97} presented a classic HDRI algorithm that involves response curve fitting and radiance image fusion from multiple LDR images. Here, we adopt their algorithm to gain the inverse camera response function $g(Z)$. This curve maps from pixel value $Z$ to product of radiance $L$ and exposure time $t$. Because $Z$ is discrete (typically 0-255), they took several photos of different exposure times and managed to solve $g(Z)$'s finite values as an overdetermined linear least square problem. Since it is neither feasible nor necessary to include all pixels to compose constraint equations, they picked up some of them by hand. We instead choose a more elaborate pixel sampling strategy to eliminate effect of noise or blurring, so as to achieve more smooth and robust result. Figure~\ref{calibration} shows the idea. We generate a virtual color card - an image of 20$\times$15 randomly colored square tiles in high resolution (a convenient substitute for a real Macbeth chart). We display this image on a digital screen (like one of an iPad) and take photos. Afterwards we resize the photos to 20$\times$15 using the average filter, producing 300 samples which are resistant to undesirable artifacts that may affect single pixel. We show that the recovered curve only coincides with gamma 2.2 curve in a middle interval.
With the recovered function $g(Z)$, we apply it mapping point image to a radiance map as the target of BRDF fitting, rather than gamma correction. Extra point images can be taken and merge into one radiance map if a single shot is insufficient to cover the extremely high variance of radiance in some cases, especially for polished metals. The effect of multiple shot will be shown in Section~\ref{verification}.

\subsection{Gray Card Calibration}
\label{graycard}
To optimize $\mathbf{x}(\mathbf{p})$ in Equation~\ref{render-model}, we need to know point light intensity $E$. $E$ can be arbitrarily assigned but lose the generality of real world physical properties. To solve this problem, we use a point image of a material with known BRDF and reversely solve $E$ first. A gray card commonly used in photography seems an ideal choice, for it can be seen as a simple Lambertian object with 18\% reflectance across the spectrum, and cheap to obtain as well. We take a photo of the gray card under mobile phone's flash (see Figure~\ref{calibration}), and record the perpendicular distance $r_{\perp}$ from phone plane to material plane (rounded to 0.5cm). In Equation~\ref{render-model}, gray card has $f = \frac{0.18}{\pi}$ that is independent of light, view direction or surface position. Now that we know $L$ is the radiance map of gray card photo, light intensity can be easily computed by pixel-wise division and then averaged, noted as $E_{gray}$.

When shooting the point image of a material sample, we keep ISO, lens aperture (often fixed on mobile device) and white balance settings coincident with gray card image. Perpendicular distance $r_\perp$ and exposure time may vary, so $E$ in each material sample is derived from $E_{gray}$ proportional to exposure time and inverse square of $r_\perp$. With gray card calibration, the reflectance properties of target materials are bound to our gray card with known absolute reflectance, rather than lighting or camera setting conditions. In other words, we measure material appearance with a gray card as reference, $E$ just becomes a mediate variable.

Response curve recovering and gray card calibration are specific to imaging device and lighting, so they are only needed once for a specific mobile phone.

\subsection{Image Alignment}
\label{alignment}
We recommend to use a tripod to take experiment photos, but there are situations that slight motions may occur when operating on smartphone's touchscreen or setting up the tripod is inconvenient. During response curve recovery, we wish to align a stack of color card photos with different exposures. Since these images are downsampled and the number of photos is relatively large, translation motion model is adequate. We use median threshold bitmap (MTB) algorithm~\cite{War03} to align them. As for per material sample images, we select a point image as target and the rest (if any) point images and the ambient image are transformed using enhanced correlation coefficient (ECC) maximization algorithm~\cite{EP08}. ECC features homography transformation and pixel-level precise alignment is achieved, so that clustering info in ambient image (see Section~\ref{clustering}) matches pixels in point image(s). We employ MTB and ECC implementations in OpenCV~\cite{Bra00}.

\section{Fitting BRDF}
\label{fitting-brdf}
In this section, we narrate our algorithm for fitting BRDF and bump maps using an ambient image and a point radiance map. First we perform clustering on the ambient map to reduce the number of spatially varying variables to number of clusters. Different pixels within a cluster can be regarded as observed samples at different positions. The main challenge we are facing at this stage is the \emph{ambiguity} among parameters. For instance, if the observed radiance is dark, it may be explained that the albedo (\textit{baseColor}) is small, or the normal's orientation is away from light source, or adjusting \textit{roughness} to get similar results. This problem is limited by the fact that a cluster contains many pixels as constraints, but still requires careful treatment. Hence we make use of an iterative multi-stage process to accomplish BRDF fitting:
\begin{algorithm}
      \SetKwFor{RepTimes}{repeat}{times}{end}
      Partition pixels into $k$ clusters\;
      \RepTimes{n}{
            Compute height map and derive normal map\;
            Optimize global BRDF parameters\;
            Optimize svBRDF parameters for each cluster\;
            Apply Gaussian blurring to svBRDF maps\;
      }
\end{algorithm}

In each iteration, quantities to be solved are separated into three parts and treated differently to suppress ambiguity. The height map is derived from the ambient image and \textit{baseColor} map; \textit{roughness} and \textit{metallic} are fitted as global parameters; and the rest svBRDF parameters are solved per cluster. These three steps are arranged in the specified order: height map relies on \textit{baseColor} of last iteration; global parameters account for overall highlight distribution pattern; and finally spatially varying ones are fitted to incarnate details and minimize the difference between rendered image and real photo. For logical continuity, narrative order is not the same as implementation order. For those spatially varying quantities, we maintain two data structures that store values per pixel (i.e., a map) and per cluster respectively. Finally, at the end of each iteration, we apply Gaussian blurring to spatially varying data maps and the average value in each cluster is used as initial guess of per cluster optimization in next iteration. We set up $n=5$ iterations. This iterative strategy refines rough results gradually to precise ones, for we exponentially decrease Gaussian blurring standard deviation and stopping criteria (relative error tolerance) of numerical optimization.

\subsection{Clustering Pixels}
\label{clustering}
We facilitate the problem by applying the concept of self-similarity: the spatially varying material is composed of a limited set of materials. By clustering pixels into a small number of sets $\{C_j\},~j=1,...,k$, we only need to fit BRDF parameters for each set.

A simple way of clustering pixels is to consider classification only according to pixel colors, which shares an identical idea with color quantization. To further consider neighborhood structures of the material and make clustering results more resistant to uneven illumination, we also use BRIEF descriptor~\cite{CLSF10} to produce a binary vector describing one pixel's local feature. This descriptor is categorical of two categories 0/1 with number of dimensions equal to length of bits in the descriptor. We use k-prototypes, an analog of k-means algorithm to cluster on the mixed-type data~\cite{Hua97, Hua98}. For each pixel, 3 numerical values of pixel color $\mathbf{\rho}_{num}$ and a categorical binary descriptor $\mathbf{\rho}_{cat}$ constitute its feature vector $\mathbf{\rho}$. Distance between two pixels is defined by
\begin{equation}
      \label{kproto}
      \dist(\mathbf{\rho}^{(1)}, \mathbf{\rho}^{(2)}) = \norm{ \mathbf{\rho}^{(1)}_{num} - \mathbf{\rho}^{(2)}_{num} }^2 + \gamma \norm{  \mathbf{\rho}^{(1)}_{cat} - \mathbf{\rho}^{(2)}_{cat} }_H  
\end{equation}
where $\norm{\sbullet}$, $\norm{\sbullet}_H$ measures Euclidean and Hamming distance respectively. $\gamma$ is a weighing factor introduced to favor either type of features, and the average standard deviation of numerical attributes divided by length of BRIEF bit vector is set as default, noted as $\gamma_0$ (see discussion in Section~\ref{tweaking}). The clustering process resembles standard k-means implementation, involving iteratively assign each pixel to the cluster with nearest center and update all centers. We use k-means++ initialization method~\cite{AV07}.

Clustering is done on the ambient image. Principal component analysis~\cite{WEG87} is performed to transform RGB pixels into three independent color channels and we only compute BRIEF features on the first principle channel, as RGB values are usually correlated, leading to redundant information about local structures. To take multi-scale structure into account, bit lengths of 48, 80, 32, window sizes of 33, 17, 5 and Gaussian blur standard deviations of 4, 2, 0 are adopted respectively, as done by~\cite{AWL15} in their pixel matching step. With respect to number of clusters, 500 is appropriate in most cases that maintains details of materials and does not cause too long processing time.

\subsection{Optimizing Global Parameters}
\label{global}
\textit{roughness} and \textit{metallic} have defining roles in the properties of BRDF than other parameters (see Section~\ref{sv-global}) and are treated as global quantities. \textit{roughness} is crucial for the shape of specular reflection lobe, which indicates that the probability distribution of energy matters rather than absolute reflectance strength. We convert point radiance map to grayscale and normalize it to be summed to 1, named $\hat{L}_0$ which is a discrete probability distribution about pixel location $\mathbf{p}$. Similarly, the rendered radiance map is also converted to $\hat{L}$. We minimize the cross entropy of $\hat{L}_0$ and $\hat{L}$, to expect that $\hat{L}$ matches real distribution:
\begin{equation}
      R_{global}  = -\sum_{\mathbf{p}} \hat{L}_0(\mathbf{p}) \log \hat{L}(\mathbf{p}) 
\end{equation}
We solve this optimization problem using L-BFGS-B algorithm~\cite{BLNZ95, ZBLN97}. 
 
In practice, a value of \textit{metallic} between 0 and 1 is rarely used. So we just tag the material by hand telling if it is metal and set \textit{metallic} to 0 or 1.

\subsection{Optimizing Spatially Varying Parameters}
\label{sv}
Now that global parameters have determined the approximate shape of BRDF lobe, spatially ones (per cluster) are responsible for details and absolute matching between the rendered image and photo. These parameters are \textit{baseColor}, \textit{specular}, \textit{specularTint} and \textit{anisotropic}. Besides them, tangent $\mathbf{t}$ is also solved together because it is related to anisotropy. With normal $\mathbf{n}$ settled at the beginning of each iteration (discussed later), only one degree of freedom is left to determine $\mathbf{t}$ because $\mathbf{n}$ and $\mathbf{t}$ are perpendicular. We parameterize azimuthal angle $\phi_t$ with another 0-1 bounded variable \textit{anisoAxis}, here $\phi_t = anisoAxis \cdot 2 \pi$. Now that $n=[n_x, n_y, n_z]$ is known and $\mathbf{n}\cdot\mathbf{t}=0$, where $t=[\sin\theta_t\cos\phi_t, \sin\theta_t\sin\phi_t, \cos\theta_t]$, substitute them into equation and zenithal angle $\theta_t$ can be determined, and thus $\mathbf{t}$:
\begin{equation}
\begin{aligned}
      \sin\theta_t &= \frac{n_z}{\sqrt{(n_x\cos\phi_t+n_y\sin\phi_t)^2+n_z^2}} \\
      \cos\theta_t &= \frac{-(n_x\cos\phi_t+n_y\sin\phi_t)}{\sqrt{(n_x\cos\phi_t+n_y\sin\phi_t)^2+n_z^2}}
\end{aligned}
\end{equation}

$t$ is only effective in anisotropic models that controls the direction of anisotropy while \textit{anisotropic} controls strength. If half vector (bisection of directions towards light and eye) rotates in the plane perpendicular to $n$, when aligned with $t$, the normal distribution function (NDF) in microfacet model has maximum response. Now we let \textit{anisoAxis} and the other four parameters to minimize the average difference between the rendered pixels for each cluster:
\begin{equation}
      R_{cluster(j)} = \frac{1}{|C_j|} \sum_{\mathbf{p}\in C_j} |L(\mathbf{p}) - L_0(\mathbf{p})|
\end{equation}
In implementation, Pseudo-Huber loss can be applied to the fitting residual $|L(\mathbf{p}) - L_0(\mathbf{p})|$ to approach better continuity near zero.

\subsection{Compute Height Map and Normals}
\label{heightmap}
We compute normals in a different way from optimizing other parameters, in order to limit degrees of freedom of numerical optimization. With an estimated \textit{baseColor} map from fitting process in previous iteration, we try to derive height map from the ambient image using an algorithm proposed by \cite{GWM*08}. The computation is efficient and pixel-wise, producing detailed height maps. If we regard \textit{baseColor} as diffuse albedo, shading image $S$ can be computed simply dividing ambient image with \textit{baseColor} in grayscale space, and it is normalized to have mean pixel intensity 0.5 corresponding to depth value 1. Inspired by idea of dark-is-deep, they treat valleys and hills on flat surface as pits of cylinders and protrusions of hemispheres, effectively forming an analytical curve which maps $S$ to depth for each pixel:
\begin{equation}
      D(S) = 
      \begin{cases}
        \sqrt{1/S - 1} & S \leq 1/2 \\
        2(1-S) & S > 1/2
      \end{cases}
\end{equation}

To recover height (depth) map at multiple scales, $S$ is Gaussian filtered with $N$ standard deviations $r_i$ in ascending order, forming $S_i$. The curve is applied on incremental shading image $l_i = 0.5S_i/S_{i+1}$ on the first $N-1$ level and $l_N = S_N$ as basal level, then accumulated to a depth map:
\begin{equation}
      \label{sigma}
      d = \sigma\sum_{i=1}^N r_i (D(l_i)-1)
\end{equation}
$l_i$ is calculated that preserves mean intensity of 0.5, and $D(l_i)$ is subtracted by 1 to make zero average depth on each level so that depths are summable among all levels. We pick four levels with Gaussian deviations 1, 2, 4 and 8. Inferring $\mathbf{n}$ from depth $d$ is a process of finding normals of a 3D isosurface:
\begin{equation}
\begin{aligned}
      \tilde{\mathbf{n}} &= [\frac{\partial d}{\partial x}, \frac{\partial d}{\partial y}, 1] \\
      \mathbf{n} &= \frac{\tilde{\mathbf{n}}}{\norm{\tilde{\mathbf{n}}}} 
\end{aligned}
\end{equation}
gradient vectors are calculated by Sobel operators. Before converting to gradient, the depth map is multiplied by a scaling factor $\sigma$ to control the overall "strength" of the normal map, its default value is 0.5 (see Section~\ref{tweaking} for discussion). 

Note that during the first iteration the ambient image and \textit{baseColor} map are identical, so all $\mathbf{n}$ naturally face upward.

\section{Result and Evaluation}
\label{experiment-analysis}
\subsection{Implementation and Recovering Result}
Our pipeline is implemented in pure Python and relies heavily on NumPy~\cite{vdWCV11} and SciPy~\cite{JOP*01} for efficient array computation and numerical optimization. Processing on one material sample usually takes three to four hours on a server with an Intel Xeon E5-2697 CPU. At present, some parts of the code such as k-prototypes clustering remains unoptimized and can be parallelized on multi-core CPU or GPU, thus there is still potential for performance boost.

\begin{figure*}[tbp]
      \centering
      \includegraphics[width=1.0\linewidth]{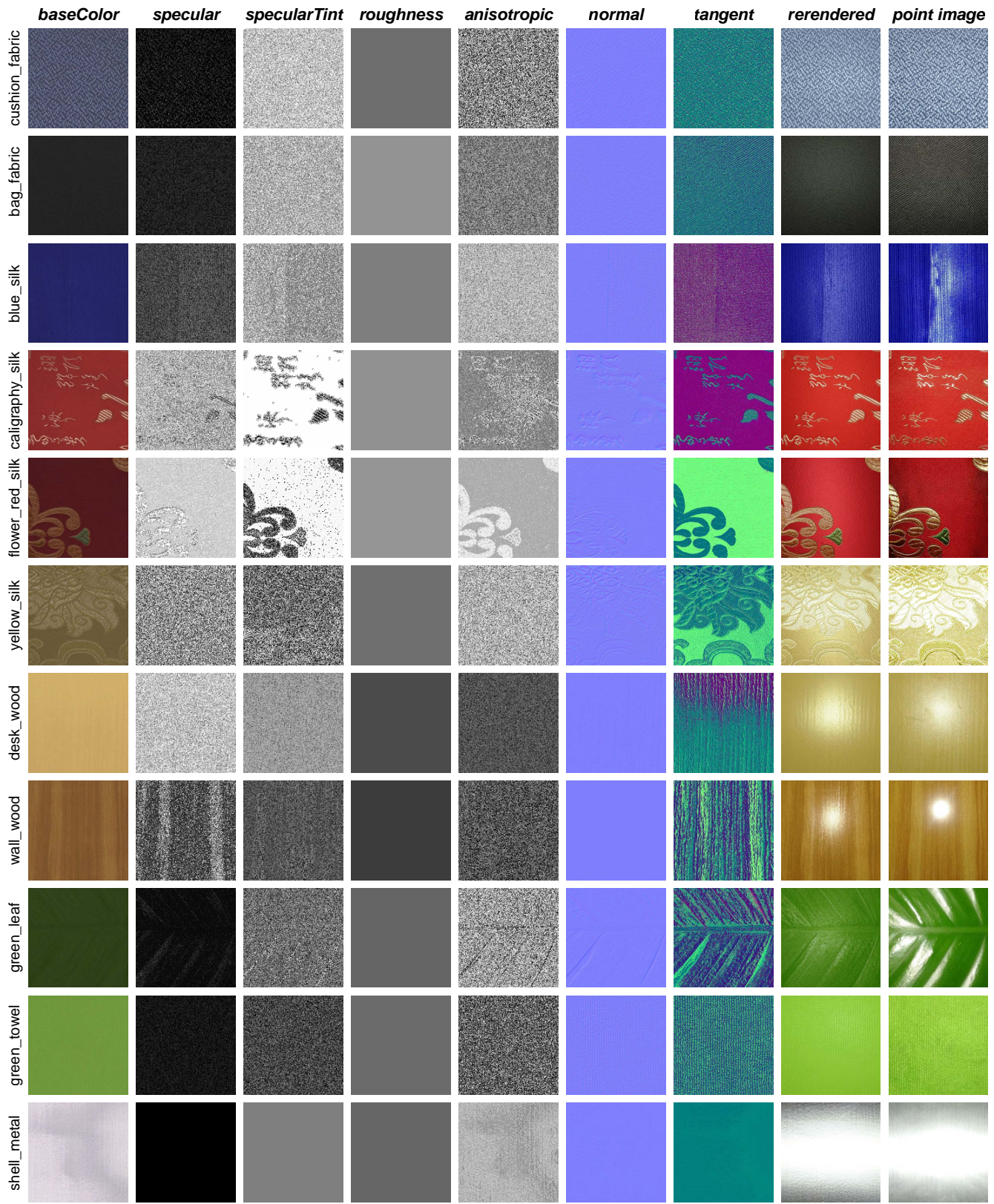}
      %

      \caption{\label{pairwise}
               Results of svBRDF fitting on our dataset. For better comparison, the rendered radiance $L$ is multiplied by exposure time and mapped to LDR images using inverse function of recovered $g(Z)$. \textit{metallic} is 0 except \textit{shell\_metal} and not shown in the figure.}
\end{figure*}

We capture the experiment photos with a HUAWEI Honor 9 which features photo resolution of 3968$\times$2976. We use its built-in camera application and manually set color temperature, ISO and exposure time. It is key to keep ISO and color temperature unchanged when taking gray card image, color card images and point images for all materials, so that camera response curve remains consistent and light intensity $E$ can be scaled according to $r_{\perp}$ and exposure time. We present a selection of results side by side in Figure~\ref{pairwise}.

\subsection{Verification}
\label{verification}
\begin{figure*}[tbp]
      \centering
      \includegraphics[width=1.0\linewidth]{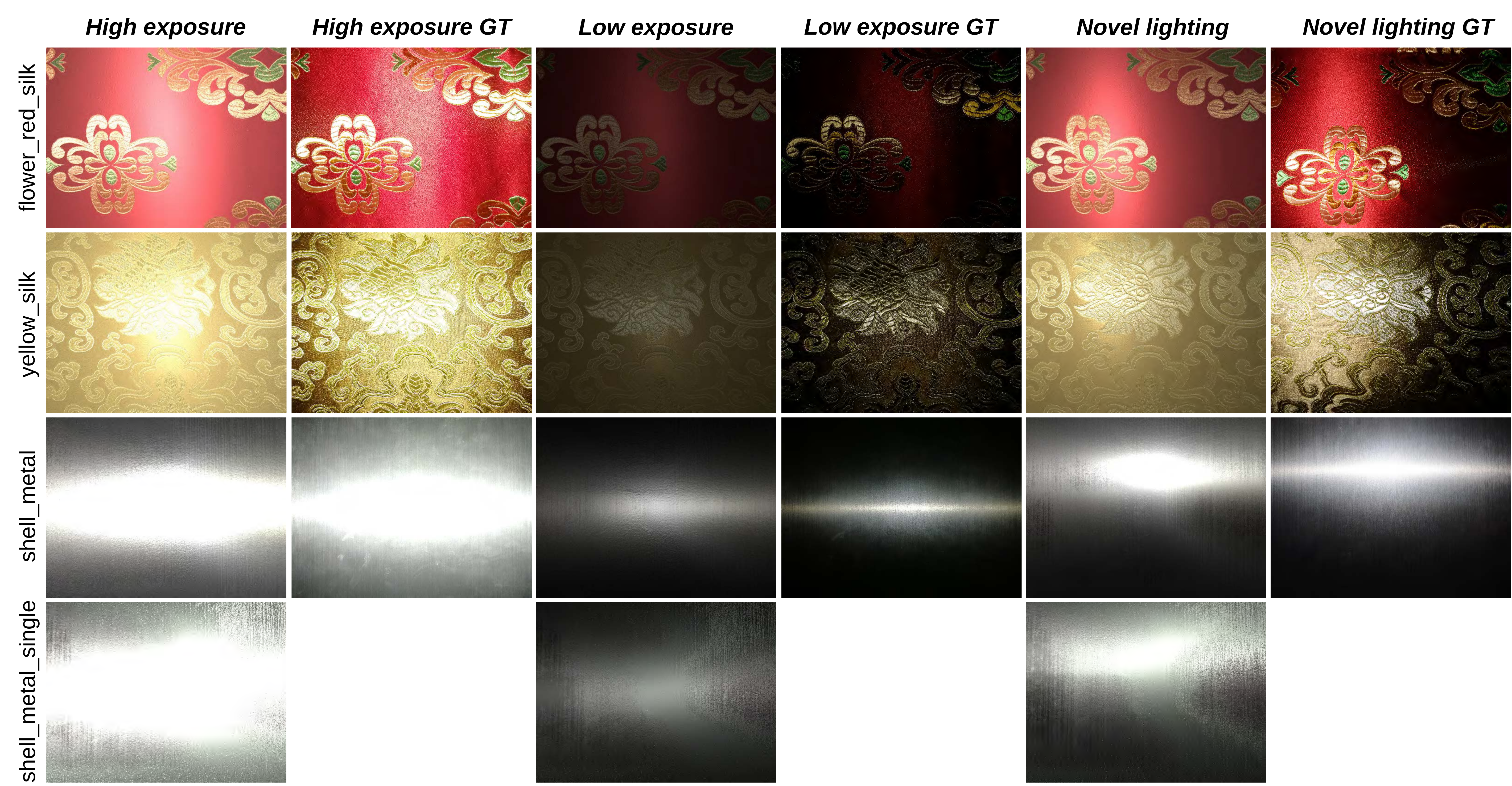}
      \caption{\label{novel}
               Rendering under novel conditions. \textit{shell\_metal\_single} is produced using the same input as \textit{shell\_metal}, with difference that only one point image of three is used for radiance map composition. }
\end{figure*}

We demonstrate our pipeline produces svBRDF textures that are faithful for photorealistic image synthesis. We render images under short and long exposure times, and a new lighting position. They are listed in Figure~\ref{novel} along with ground-truth photos. We also show that taking multiple point images helps restoring details under low or high exposure. These images are merged to a more accurate radiance map when a single LDR image cannot hold sufficient information. A video is attached in the supplemental material for better showcase. 

\begin{figure}[htb]
      \centering
      \includegraphics[width=0.9\linewidth]{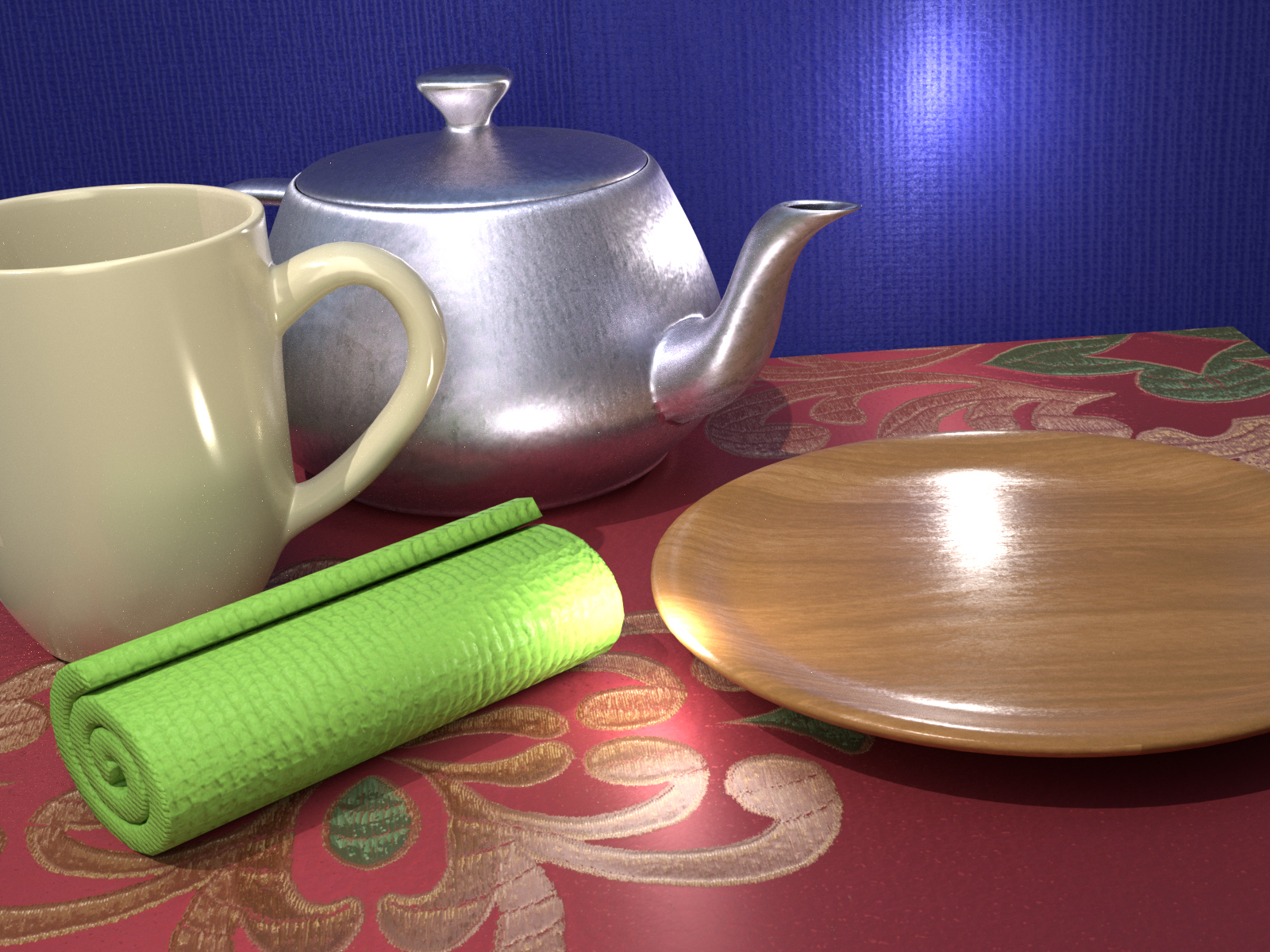}
      \caption{\label{scene}
            A scene showing textured 3D models with svBRDF maps produced by our pipeline. }
\end{figure}

We also map the svBRDF textures onto objects and render a photorealistic scene (Figure~\ref{scene}), using an open source 3D software Blender and its physically based renderer Cycles (\url{https://www.blender.org/}). The scene features materials of metal, ceramic, silk, wood and cotton illuminated by an HDR environment image. Some point lights are also added to demonstrate specular and anisotropic properties of materials.

\subsection{Comparisons}
\label{comparison}
\begin{figure*}[tbp]
      \centering
      \includegraphics[width=0.9\linewidth]{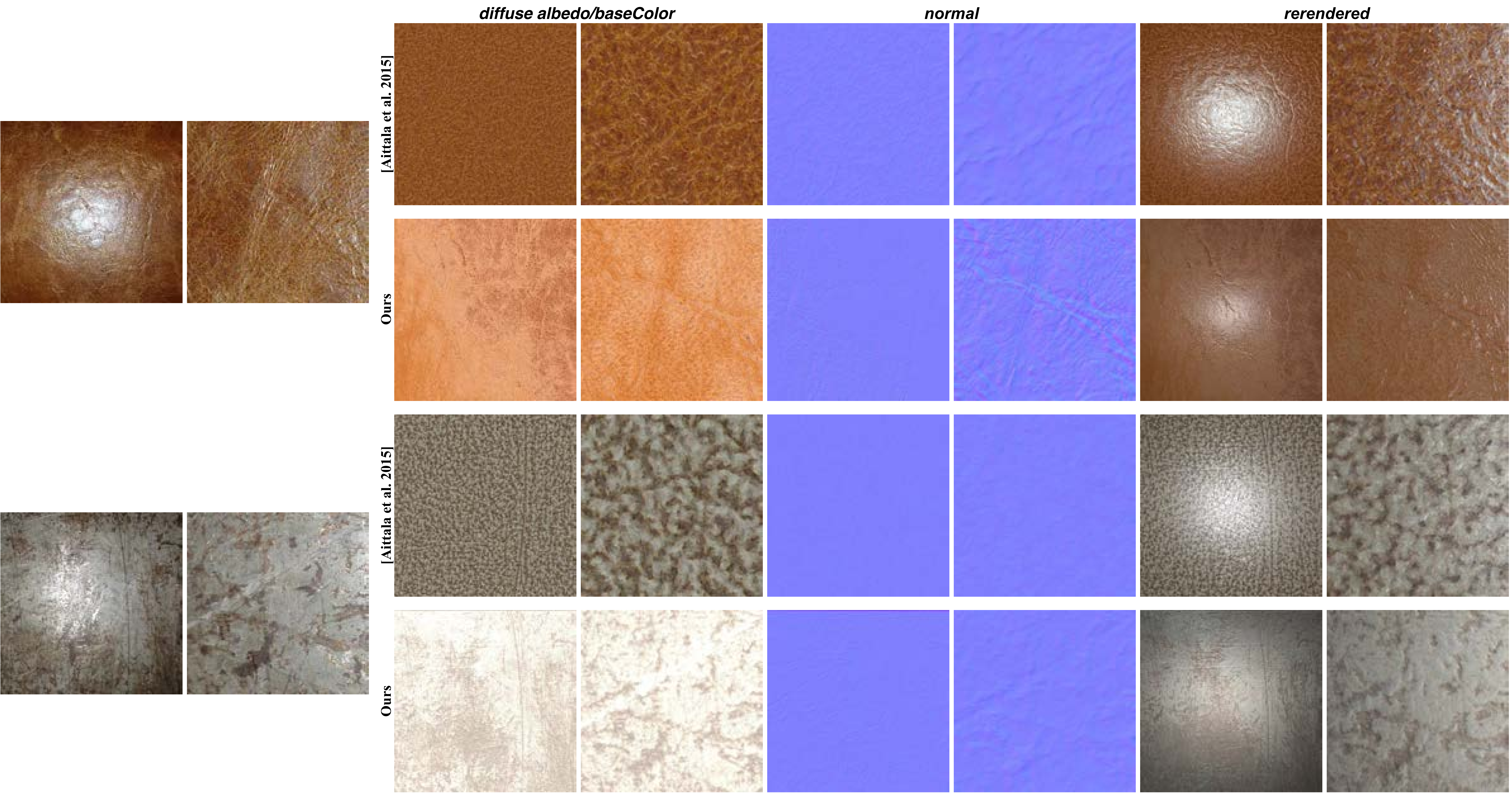}
      \caption{\label{ai}
            Comparison with method by Aittala et al.~\cite{AWL15}. We show the input image under flash (with close-up) with one recovered parameter (\textit{baseColor} in our method, diffuse albedo in theirs), normal and rerendered result. Note that the parameter is not directly comparable with respect to absolute values due to different BRDF models and arbitrarily set light intensity. }
\end{figure*}

Our work shares basic idea with the method proposed by Aittala et al.~\cite{AWL15}.  We compare some results of our pipeline to theirs, with input images they supplied (Figure~\ref{ai}, some of their input photos are also tested in subsequent experiments, though lacking of calibration images). Both methods take two photos under ambient (not shown in the figure) and flash light as input and can produce high-resolution textures. A big difference is that they cut the image into several tiles and find relations between one chosen representative tile (\textit{master tile}) and others. This impose a restriction that all tiles must have similar structural compositions (that is to say, the whole material is "texture-like"), while this is not the case in many situations. Their deliberately optimization process involves complex regulation terms, which successfully captures visual properties of materials. However, the structures of generated textures do not match the original very well when input images show irregular patterns, yet our clustering-based method is unrestricted.

We also compare our work with the method by~\cite{DAD*18}, which is representative among recent deep learning based appearance modeling approaches. Their proposed Convolutional Neural Network (CNN) only requires one image lit by a flash light. But the input and results are limited to resolution of 256$\times$256, which causes much loss of details. Their method fails when the input photo exhibits strong specular highlight (see discussion in Section~\ref{sv-global} for explanation of our choice to not estimate \textit{roughness} as global).

\begin{figure}[htb]
      \centering
      \includegraphics[width=0.9\linewidth]{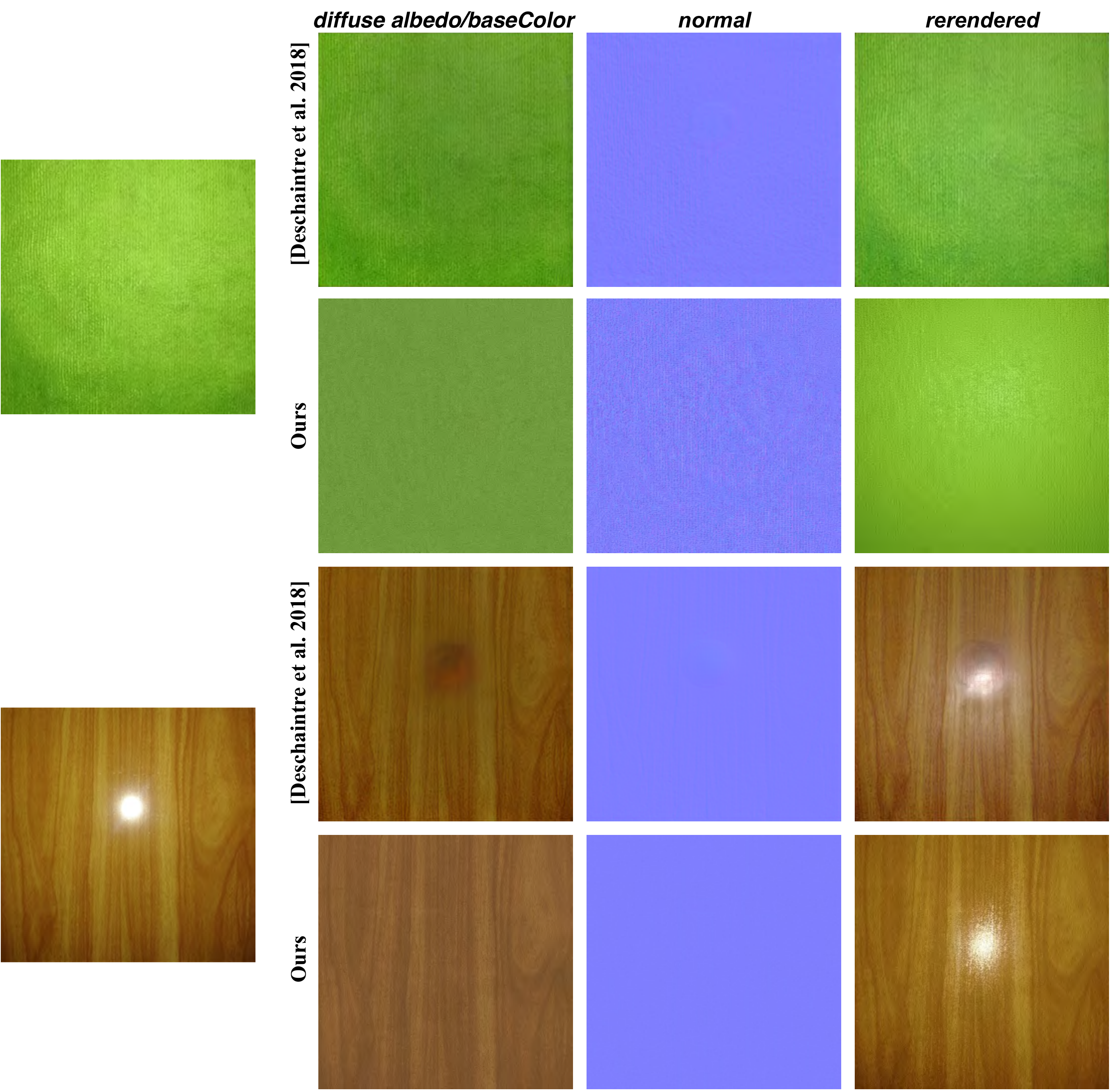}
      \caption{\label{deep}
        Comparison with method by Deschaintre et al.~\cite{DAD*18}. We feed their trained CNN with our captured images, cropped and resized as demand. The second example demonstrates a typical failure case of their method. }
\end{figure}

\section{Discussion}
\subsection{Treating BRDF Parameters Differently}
\label{sv-global}
\begin{figure}[htb]
      \centering

      \centering
      \begin{subfigure}{0.8\linewidth}
            \centering
            \includegraphics[width=\linewidth]{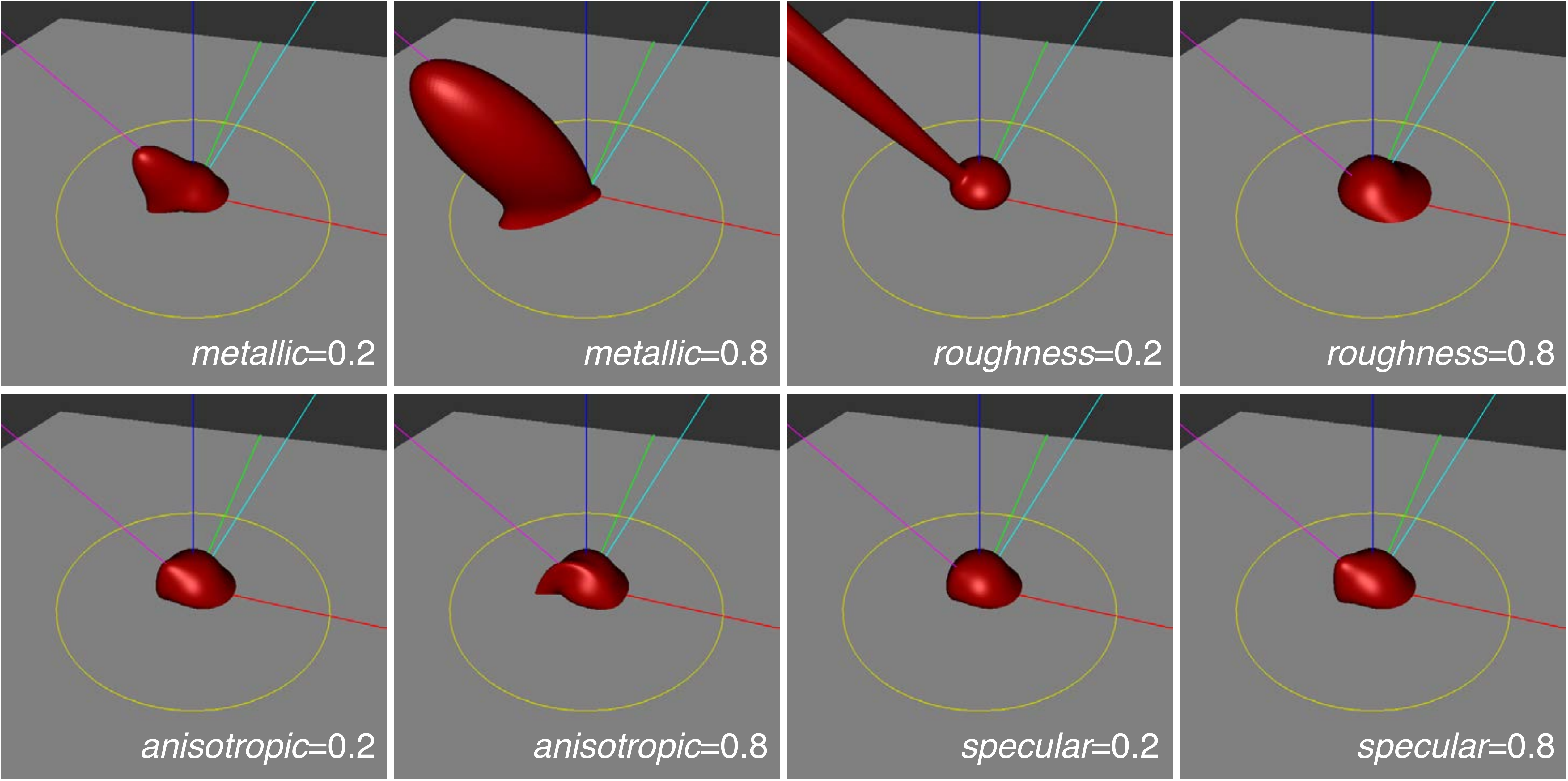}
            \caption{}
      \end{subfigure}
      \begin{subfigure}{0.8\linewidth}
            \centering
            \includegraphics[width=\linewidth]{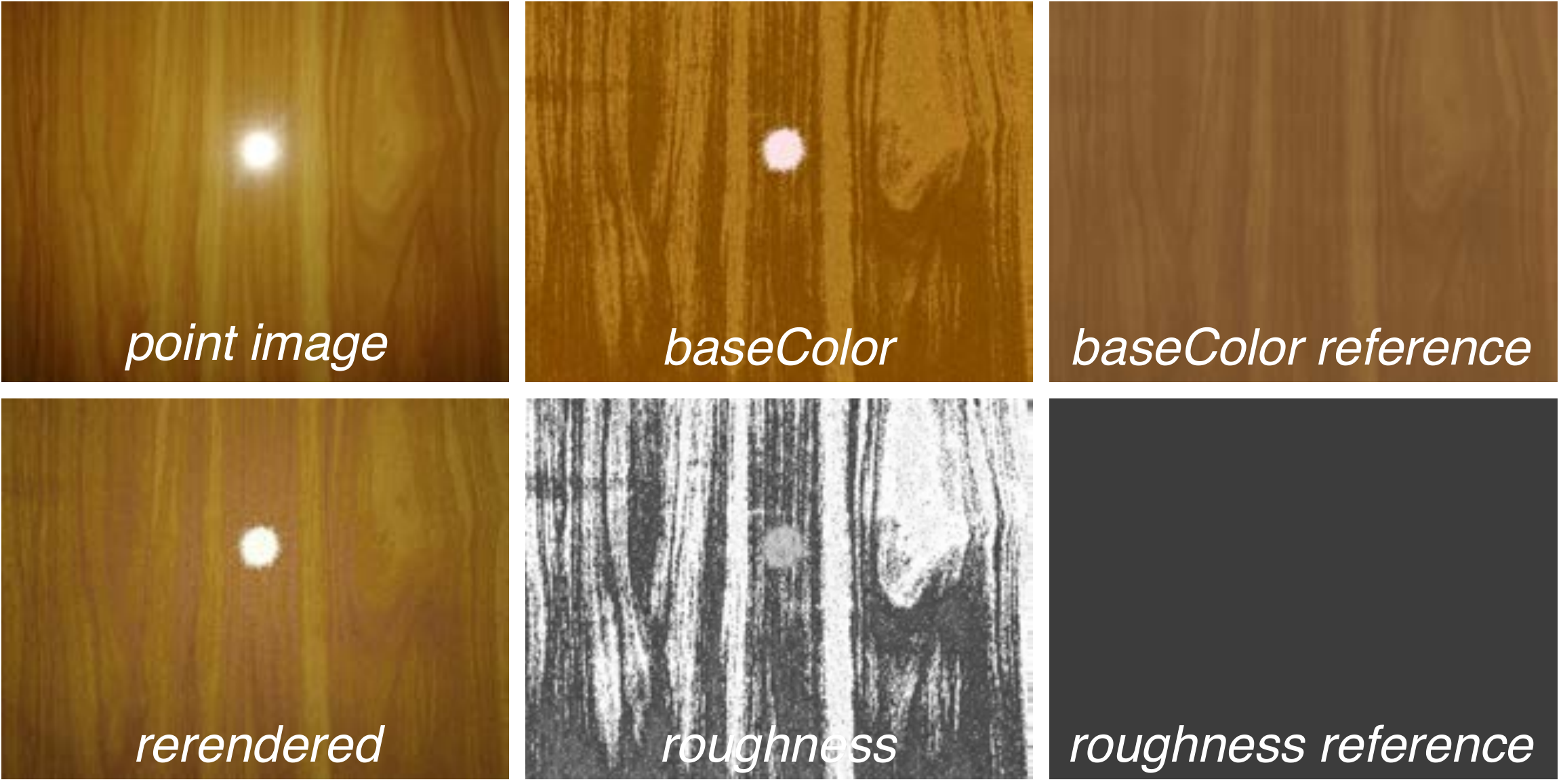}
            \caption{}
      \end{subfigure}

      \caption{\label{global-affect}
          (a) 3D plots of Disney BRDF lobes in the BRDF Explorer~\cite{Bur12}. Parameters are set as default except ones noted in the figures. \textit{roughness} and \textit{metallic} make decisive differences while other parameters just affect BRDF in a relatively local and restricted manner. (b) Artifacts appear when setting \textit{roughness} as a spatially varying parameter. The highlight in point image is not removed from solution maps (middle column), and rendering will be incorrect using novel light or view position. We attach the result keeping \textit{roughness} global as reference in right column. }
\end{figure}

Parameters listed in Section~\ref{render} present different properties. We observed that \textit{roughness} and \textit{metallic} dramatically affect the shape of BRDF lobes compared to the others (Figure~\ref{global-affect}). Our experiment shows that if \textit{roughness} is spatially varying, undesirable artifacts will occur in the result. Because the value of radiance inside highlight spot is extremely large compared to the other regions, a fixed bright spot shows up in \textit{baseColor} map, giving a wrong explanation of high albedo rather than specular reflection, which "overfits" the point image. Since \textit{roughness} greatly affects the overall highlight distribution, it is solved alone as a global parameter. As for \textit{metallic}, the  and the metallic model are markedly different (metallic model has no diffuse lobe at all).

\subsection{Effect of Pipeline Parameter Tuning}
\label{tweaking}
\begin{figure}[htb]
      \centering
      \includegraphics[width=0.9\linewidth]{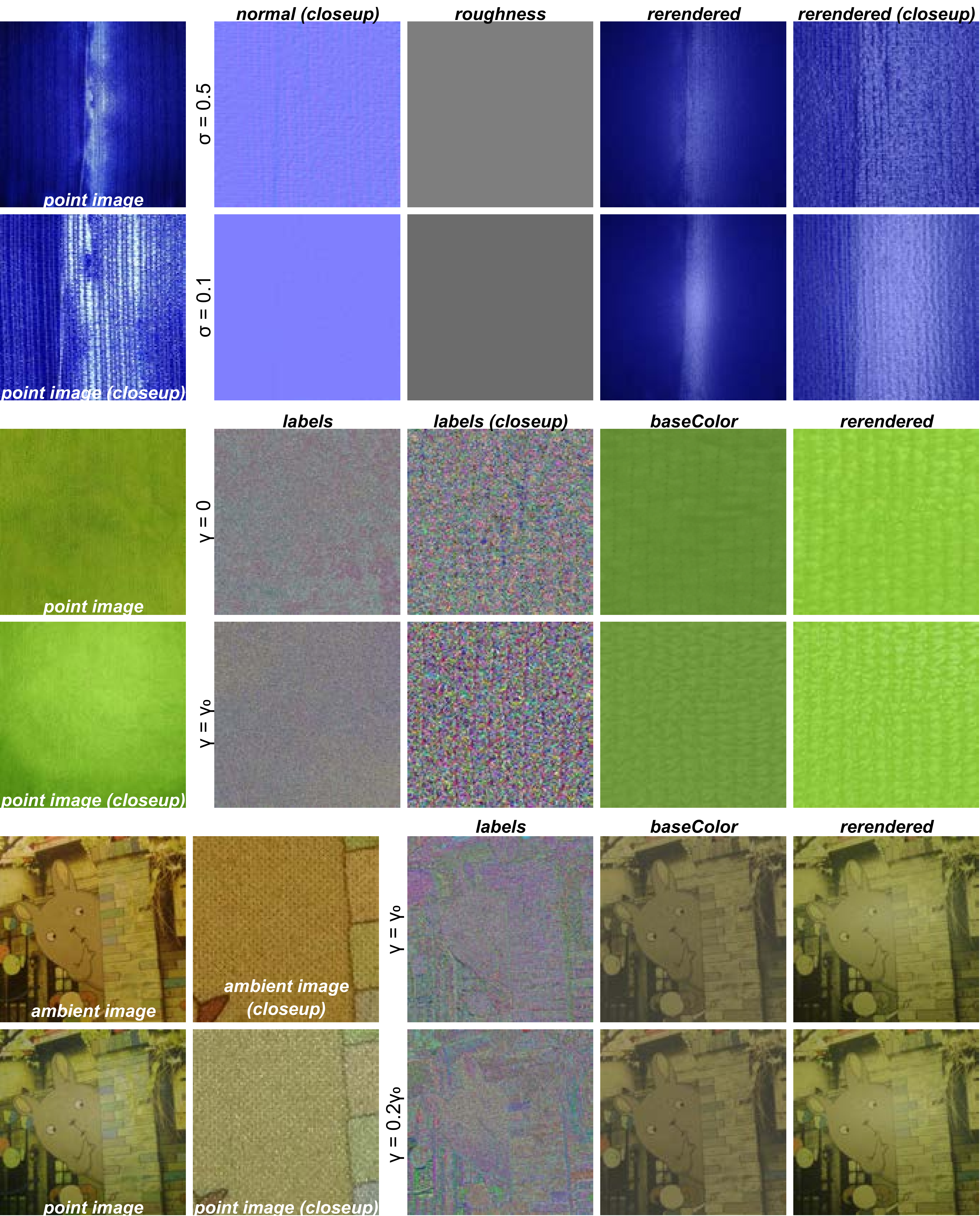}
      \caption{\label{tuning}
        Tuning of some parameters of the pipeline can ameliorate visual appearances of results. Top: with a smaller value of $\sigma$, the rerendered point image coincides with the original better, as the actual target material is fairly smooth. Middle: setting $\gamma$ to 0 makes the clustering algorithm degenerates into simple form of k-means, which makes the clustering result more noise-like (in image "labels", each color signify one kind of cluster). In case of clustering on ambient image with significant shadows or noise, this may lead to artifacts. Bottom: in some rare cases, categorical features are overestimated and the color variations are vague in generated textures. }
\end{figure}

Some parameters in our pipeline are needed to be set empirically, which may influence the appearance of results (Figure~\ref{tuning}). First of all, because it is hard to know whether the luminance variations on the material are consequence of geometry or reflectance properties, thus the factor $\sigma$ in Equation~\ref{sigma} for scaling height map preserves some freedom for tuning between them. Usually, default $\sigma$ of 0.5 is good under most circumstances, yet a smaller value is preferred for those smooth materials.

Choice of weighing factor $\gamma$ for our hybrid clustering (see Section~\ref{clustering}) is also important. We demonstrate that with categorical BRIEF features, the clustering process can produce structural clustering patterns with less noise (Figure~\ref{tuning} middle). However, colors of pixels should also not be neglected as well. In the last example of Figure~\ref{tuning}, color variations are not properly shown in the result, due to the local structures of the fabric mousepad do not coincide with surface colors. The result can be revised by scaling $\gamma$ with 0.2 to emphasize colors over BRIEF features during clustering. These empirical choices could be made with extra care when it is desirable to augment the quality of some individual cases, while default values usually seem reasonable.

\subsection{Limitations}
\label{global-roughness}
\begin{figure}[htb]
      \centering
      \begin{subfigure}{0.90\linewidth}
            \centering
            \includegraphics[width=\linewidth]{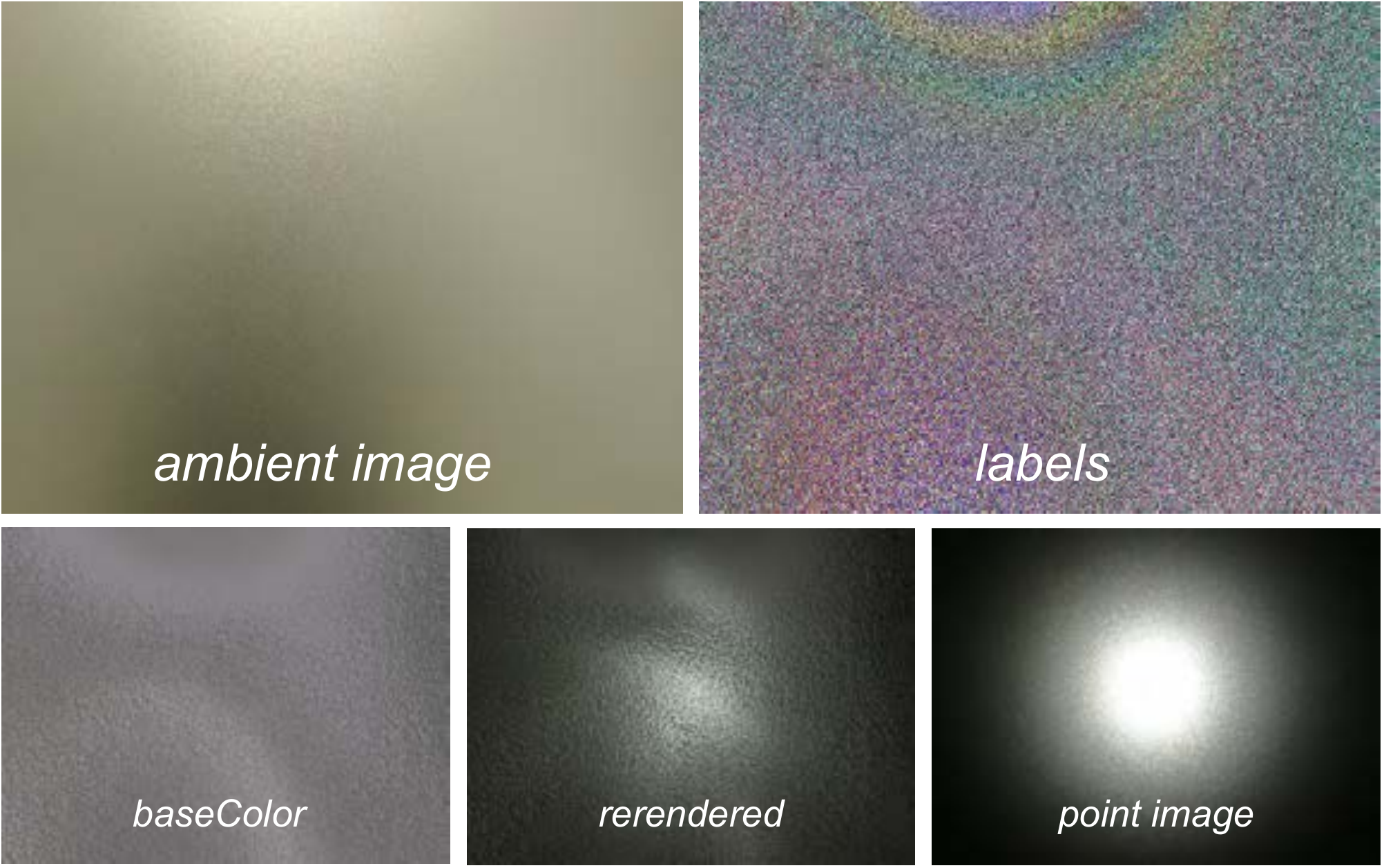}
            \caption{}
      \end{subfigure}
      \begin{subfigure}{0.90\linewidth}
            \centering
            \includegraphics[width=\linewidth]{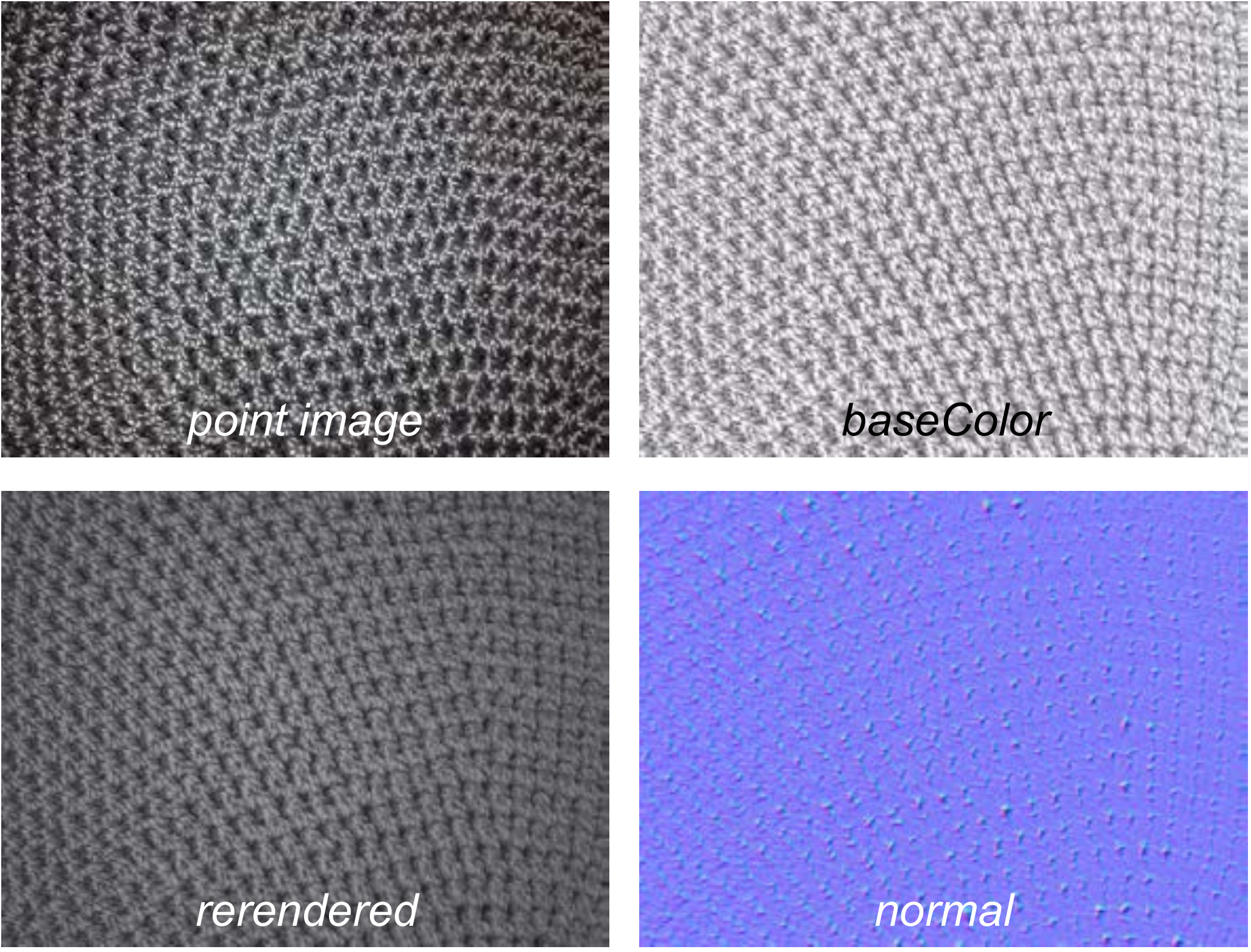}
            \caption{}
      \end{subfigure}
      \caption{\label{failure}
       Failure examples in our method. (a) Artifacts caused by inaccurate clustering. The material exhibits few structural features such that BRIEF feature detector provides poor information to aid clustering, thus shadows in the ambient image can result in artifacts. (b) The method proposed by~\cite{GWM*08} for computing height map in our pipeline cannot deal with very complex geometry. The underlying model assumes that bumps on surface can be seen as cylindrical and spherical, while this is not the case in (b). The material is woven by fabrics with strong anisotropy. However, the specular highlight is mistakenly explained in \textit{baseColor} due to falsely computed geometry.   }
\end{figure}

Failure cases occur when the ambient image provide insufficient or erroneous information that misleads clustering (Figure~\ref{failure}). If a material presents merely spatially varying patterns, the result is susceptible to shadows in the ambient image. Considering the near distance between the material and mobile phone (typically 20cm), shadowing is difficult to avoid when taking photos under ambient light, so eliminating the influence of shadows will be desirable in future work. Another problem is that the implemented height map estimating method in Section~\ref{heightmap} produces incorrect normals, when the surface is extremely bumped that deviates assumption of being nearly planar. Input images in figure~\ref{failure} (b) are from the dataset provided by Aittala et al.~\cite{AWL15}. Their method solves normals as part of optimization, which performs well in similar cases with complex geometry.

Treating \textit{roughness} and \textit{metallic} as global parameters introduces oversimplified assumptions, but is necessary for faithful outputs as shown in Figure~\ref{global-affect}. Nonetheless, regarding \textit{roughness} as global does impose an excessive restrction and causes loss of detail in some degree. Sophisticated clustering ideas such as hierarchical levels of details may be added, so that those decisive parameters can be fitted at a coarser level. Furthermore, our current clustering method is unable to realize superfine details. For example, \textit{shell\_metal} in Figure~\ref{pairwise} has thin brushed strips on surface, but the recovered one has more granular patterns. Instead of detecting BRIEF features at some predefined scales (Section~\ref{clustering}), adaptive algorithms should be considered to model both coarse and subtle spatial variations in textures.

\section{Conclusion}
We present a pipeline for modeling svBRDF parameters with a minimum of two smartphone photos for each planar material sample. We introduce a mixed-type feature vector for pixel clustering and a multi-stage iterative optimization process to fit parameters. The image-based calibration method can help decide unknown camera response curve and light source intensity. By testing our algorithm on a variety of materials and comparing with previous work, we demonstrate that our pipeline strikes a good balance between input complexity and result fidelity, hence becomes a novel solution to appearance modeling for both research and application purposes.

\bibliographystyle{ACM-Reference-Format}
\bibliography{references}       


\begin{thebibliography}{00}


\ifx \showCODEN    \undefined \def \showCODEN     #1{\unskip}     \fi
\ifx \showDOI      \undefined \def \showDOI       #1{#1}\fi
\ifx \showISBNx    \undefined \def \showISBNx     #1{\unskip}     \fi
\ifx \showISBNxiii \undefined \def \showISBNxiii  #1{\unskip}     \fi
\ifx \showISSN     \undefined \def \showISSN      #1{\unskip}     \fi
\ifx \showLCCN     \undefined \def \showLCCN      #1{\unskip}     \fi
\ifx \shownote     \undefined \def \shownote      #1{#1}          \fi
\ifx \showarticletitle \undefined \def \showarticletitle #1{#1}   \fi
\ifx \showURL      \undefined \def \showURL       {\relax}        \fi
\providecommand\bibfield[2]{#2}
\providecommand\bibinfo[2]{#2}
\providecommand\natexlab[1]{#1}
\providecommand\showeprint[2][]{arXiv:#2}

\bibitem[\protect\citeauthoryear{Aittala, Aila, and Lehtinen}{Aittala
  et~al\mbox{.}}{2016}]%
        {AAL16}
\bibfield{author}{\bibinfo{person}{Miika Aittala}, \bibinfo{person}{Timo Aila},
  {and} \bibinfo{person}{Jaakko Lehtinen}.} \bibinfo{year}{2016}\natexlab{}.
\newblock \showarticletitle{Reflectance modeling by neural texture synthesis}.
\newblock \bibinfo{journal}{{\em ACM Transactions on Graphics (TOG)\/}}
  \bibinfo{volume}{35}, \bibinfo{number}{4} (\bibinfo{year}{2016}),
  \bibinfo{pages}{65}.
\newblock


\bibitem[\protect\citeauthoryear{Aittala, Weyrich, and Lehtinen}{Aittala
  et~al\mbox{.}}{2013}]%
        {AWL13}
\bibfield{author}{\bibinfo{person}{Miika Aittala}, \bibinfo{person}{Tim
  Weyrich}, {and} \bibinfo{person}{Jaakko Lehtinen}.}
  \bibinfo{year}{2013}\natexlab{}.
\newblock \showarticletitle{Practical SVBRDF capture in the frequency domain.}
\newblock \bibinfo{journal}{{\em ACM Trans. Graph.\/}} \bibinfo{volume}{32},
  \bibinfo{number}{4} (\bibinfo{year}{2013}), \bibinfo{pages}{110--1}.
\newblock


\bibitem[\protect\citeauthoryear{Aittala, Weyrich, Lehtinen,
  et~al\mbox{.}}{Aittala et~al\mbox{.}}{2015}]%
        {AWL15}
\bibfield{author}{\bibinfo{person}{Miika Aittala}, \bibinfo{person}{Tim
  Weyrich}, \bibinfo{person}{Jaakko Lehtinen}, {et~al\mbox{.}}}
  \bibinfo{year}{2015}\natexlab{}.
\newblock \showarticletitle{Two-shot SVBRDF capture for stationary materials}.
\newblock \bibinfo{journal}{{\em ACM Trans. Graph.\/}} \bibinfo{volume}{34},
  \bibinfo{number}{4} (\bibinfo{year}{2015}), \bibinfo{pages}{110--1}.
\newblock


\bibitem[\protect\citeauthoryear{Albert, Chan, Goldman, and O'Brien}{Albert
  et~al\mbox{.}}{2018}]%
        {ACGO18}
\bibfield{author}{\bibinfo{person}{Rachel~A Albert},
  \bibinfo{person}{Dorian~Yao Chan}, \bibinfo{person}{Dan~B Goldman}, {and}
  \bibinfo{person}{James~F O'Brien}.} \bibinfo{year}{2018}\natexlab{}.
\newblock \showarticletitle{Approximate svBRDF estimation from mobile phone
  video}. In \bibinfo{booktitle}{{\em Proceedings of the Eurographics Symposium
  on Rendering: Experimental Ideas \& Implementations}}. Eurographics
  Association, \bibinfo{pages}{11--22}.
\newblock


\bibitem[\protect\citeauthoryear{Arthur and Vassilvitskii}{Arthur and
  Vassilvitskii}{2007}]%
        {AV07}
\bibfield{author}{\bibinfo{person}{David Arthur} {and} \bibinfo{person}{Sergei
  Vassilvitskii}.} \bibinfo{year}{2007}\natexlab{}.
\newblock \showarticletitle{k-means++: The advantages of careful seeding}. In
  \bibinfo{booktitle}{{\em Proceedings of the eighteenth annual ACM-SIAM
  symposium on Discrete algorithms}}. Society for Industrial and Applied
  Mathematics, \bibinfo{pages}{1027--1035}.
\newblock


\bibitem[\protect\citeauthoryear{Bradski}{Bradski}{2000}]%
        {Bra00}
\bibfield{author}{\bibinfo{person}{G. Bradski}.}
  \bibinfo{year}{2000}\natexlab{}.
\newblock \showarticletitle{{The OpenCV Library}}.
\newblock \bibinfo{journal}{{\em Dr. Dobb's Journal of Software Tools\/}}
  (\bibinfo{year}{2000}).
\newblock


\bibitem[\protect\citeauthoryear{Burley}{Burley}{2012}]%
        {Bur12}
\bibfield{author}{\bibinfo{person}{Brent Burley}.}
  \bibinfo{year}{2012}\natexlab{}.
\newblock \showarticletitle{Physically-based shading at disney}. In
  \bibinfo{booktitle}{{\em ACM SIGGRAPH 2012 Courses: Practical
  physically‐based shading in film and game production}}.
  \bibinfo{publisher}{ACM}.
\newblock


\bibitem[\protect\citeauthoryear{Byrd, Lu, Nocedal, and Zhu}{Byrd
  et~al\mbox{.}}{1995}]%
        {BLNZ95}
\bibfield{author}{\bibinfo{person}{Richard~H Byrd}, \bibinfo{person}{Peihuang
  Lu}, \bibinfo{person}{Jorge Nocedal}, {and} \bibinfo{person}{Ciyou Zhu}.}
  \bibinfo{year}{1995}\natexlab{}.
\newblock \showarticletitle{A limited memory algorithm for bound constrained
  optimization}.
\newblock \bibinfo{journal}{{\em SIAM Journal on Scientific Computing\/}}
  \bibinfo{volume}{16}, \bibinfo{number}{5} (\bibinfo{year}{1995}),
  \bibinfo{pages}{1190--1208}.
\newblock


\bibitem[\protect\citeauthoryear{Calonder, Lepetit, Strecha, and Fua}{Calonder
  et~al\mbox{.}}{2010}]%
        {CLSF10}
\bibfield{author}{\bibinfo{person}{Michael Calonder}, \bibinfo{person}{Vincent
  Lepetit}, \bibinfo{person}{Christoph Strecha}, {and} \bibinfo{person}{Pascal
  Fua}.} \bibinfo{year}{2010}\natexlab{}.
\newblock \showarticletitle{Brief: Binary robust independent elementary
  features}. In \bibinfo{booktitle}{{\em European conference on computer
  vision}}. Springer, \bibinfo{pages}{778--792}.
\newblock


\bibitem[\protect\citeauthoryear{CIPA DC-008-Translation-2012}{CIPA
  DC-008-Translation-2012}{2012}]%
        {CIPA12}
CIPA DC-008-Translation-2012 \bibinfo{year}{2012}\natexlab{}.
\newblock \bibinfo{booktitle}{{\em Exchangeable image file format for digital
  still cameras: Exif Version 2.3}}.
\newblock \bibinfo{type}{Standard}. \bibinfo{institution}{Camera and Imaging
  Products Association}.
\newblock


\bibitem[\protect\citeauthoryear{CIPA DCG-001-Translation-2018}{CIPA
  DCG-001-Translation-2018}{2018}]%
        {CIPA18}
CIPA DCG-001-Translation-2018 \bibinfo{year}{2018}\natexlab{}.
\newblock \bibinfo{booktitle}{{\em Individual Guidelines for noting digital
  camera specifications on Number of pixels, Image file, and Focal length of
  the lens}}.
\newblock \bibinfo{type}{Standard}. \bibinfo{institution}{Camera and Imaging
  Products Association}.
\newblock


\bibitem[\protect\citeauthoryear{Dana, Van~Ginneken, Nayar, and
  Koenderink}{Dana et~al\mbox{.}}{1999}]%
        {DvGN*99}
\bibfield{author}{\bibinfo{person}{Kristin~J Dana}, \bibinfo{person}{Bram
  Van~Ginneken}, \bibinfo{person}{Shree~K Nayar}, {and} \bibinfo{person}{Jan~J
  Koenderink}.} \bibinfo{year}{1999}\natexlab{}.
\newblock \showarticletitle{Reflectance and texture of real-world surfaces}.
\newblock \bibinfo{journal}{{\em ACM Transactions On Graphics (TOG)\/}}
  \bibinfo{volume}{18}, \bibinfo{number}{1} (\bibinfo{year}{1999}),
  \bibinfo{pages}{1--34}.
\newblock


\bibitem[\protect\citeauthoryear{Debevec, Wenger, Tchou, Gardner, Waese, and
  Hawkins}{Debevec et~al\mbox{.}}{2002}]%
        {DWT*02}
\bibfield{author}{\bibinfo{person}{Paul Debevec}, \bibinfo{person}{Andreas
  Wenger}, \bibinfo{person}{Chris Tchou}, \bibinfo{person}{Andrew Gardner},
  \bibinfo{person}{Jamie Waese}, {and} \bibinfo{person}{Tim Hawkins}.}
  \bibinfo{year}{2002}\natexlab{}.
\newblock \showarticletitle{A lighting reproduction approach to live-action
  compositing}. In \bibinfo{booktitle}{{\em ACM Transactions on Graphics
  (TOG)}}, Vol.~\bibinfo{volume}{21}. ACM, \bibinfo{pages}{547--556}.
\newblock


\bibitem[\protect\citeauthoryear{Debevec and Malik}{Debevec and Malik}{1997}]%
        {DM97}
\bibfield{author}{\bibinfo{person}{Paul~E. Debevec} {and}
  \bibinfo{person}{Jitendra Malik}.} \bibinfo{year}{1997}\natexlab{}.
\newblock \showarticletitle{Recovering High Dynamic Range Radiance Maps from
  Photographs}. In \bibinfo{booktitle}{{\em Proceedings of the 24th Annual
  Conference on Computer Graphics and Interactive Techniques}} {\em
  (\bibinfo{series}{SIGGRAPH '97})}. \bibinfo{publisher}{ACM
  Press/Addison-Wesley Publishing Co.}, \bibinfo{address}{New York, NY, USA},
  \bibinfo{pages}{369--378}.
\newblock
\showISBNx{0-89791-896-7}
\showDOI{%
\url{https://doi.org/10.1145/258734.258884}}


\bibitem[\protect\citeauthoryear{Deschaintre, Aittala, Durand, Drettakis, and
  Bousseau}{Deschaintre et~al\mbox{.}}{2018}]%
        {DAD*18}
\bibfield{author}{\bibinfo{person}{Valentin Deschaintre},
  \bibinfo{person}{Miika Aittala}, \bibinfo{person}{Fredo Durand},
  \bibinfo{person}{George Drettakis}, {and} \bibinfo{person}{Adrien Bousseau}.}
  \bibinfo{year}{2018}\natexlab{}.
\newblock \showarticletitle{Single-image svbrdf capture with a rendering-aware
  deep network}.
\newblock \bibinfo{journal}{{\em ACM Transactions on Graphics (TOG)\/}}
  \bibinfo{volume}{37}, \bibinfo{number}{4} (\bibinfo{year}{2018}),
  \bibinfo{pages}{128}.
\newblock


\bibitem[\protect\citeauthoryear{Dong, Tong, Pellacini, and Guo}{Dong
  et~al\mbox{.}}{2011}]%
        {DTPG11}
\bibfield{author}{\bibinfo{person}{Yue Dong}, \bibinfo{person}{Xin Tong},
  \bibinfo{person}{Fabio Pellacini}, {and} \bibinfo{person}{Baining Guo}.}
  \bibinfo{year}{2011}\natexlab{}.
\newblock \showarticletitle{AppGen: interactive material modeling from a single
  image}. In \bibinfo{booktitle}{{\em ACM Transactions on Graphics (TOG)}},
  Vol.~\bibinfo{volume}{30}. ACM, \bibinfo{pages}{146}.
\newblock


\bibitem[\protect\citeauthoryear{Dong, Wang, Tong, Snyder, Lan, Ben-Ezra, and
  Guo}{Dong et~al\mbox{.}}{2010}]%
        {DWT10}
\bibfield{author}{\bibinfo{person}{Yue Dong}, \bibinfo{person}{Jiaping Wang},
  \bibinfo{person}{Xin Tong}, \bibinfo{person}{John Snyder},
  \bibinfo{person}{Yanxiang Lan}, \bibinfo{person}{Moshe Ben-Ezra}, {and}
  \bibinfo{person}{Baining Guo}.} \bibinfo{year}{2010}\natexlab{}.
\newblock \showarticletitle{Manifold bootstrapping for SVBRDF capture}.
\newblock \bibinfo{journal}{{\em ACM Transactions on Graphics (TOG)\/}}
  \bibinfo{volume}{29}, \bibinfo{number}{4} (\bibinfo{year}{2010}),
  \bibinfo{pages}{98}.
\newblock


\bibitem[\protect\citeauthoryear{Evangelidis and Psarakis}{Evangelidis and
  Psarakis}{2008}]%
        {EP08}
\bibfield{author}{\bibinfo{person}{Georgios~D Evangelidis} {and}
  \bibinfo{person}{Emmanouil~Z Psarakis}.} \bibinfo{year}{2008}\natexlab{}.
\newblock \showarticletitle{Parametric image alignment using enhanced
  correlation coefficient maximization}.
\newblock \bibinfo{journal}{{\em IEEE Transactions on Pattern Analysis and
  Machine Intelligence\/}} \bibinfo{volume}{30}, \bibinfo{number}{10}
  (\bibinfo{year}{2008}), \bibinfo{pages}{1858--1865}.
\newblock


\bibitem[\protect\citeauthoryear{Foo}{Foo}{1997}]%
        {Foo97}
\bibfield{author}{\bibinfo{person}{Sing~Choong Foo}.}
  \bibinfo{year}{1997}\natexlab{}.
\newblock {\em \bibinfo{title}{A gonioreflectometer for measuring the
  bidirectional reflectance of material for use in illumination computation}}.
\newblock \bibinfo{thesistype}{Ph.D. Dissertation}. \bibinfo{school}{Citeseer}.
\newblock


\bibitem[\protect\citeauthoryear{Glencross, Ward, Melendez, Jay, Liu, and
  Hubbold}{Glencross et~al\mbox{.}}{2008}]%
        {GWM*08}
\bibfield{author}{\bibinfo{person}{Mashhuda Glencross},
  \bibinfo{person}{Gregory~J Ward}, \bibinfo{person}{Francho Melendez},
  \bibinfo{person}{Caroline Jay}, \bibinfo{person}{Jun Liu}, {and}
  \bibinfo{person}{Roger Hubbold}.} \bibinfo{year}{2008}\natexlab{}.
\newblock \showarticletitle{A perceptually validated model for surface depth
  hallucination}. In \bibinfo{booktitle}{{\em ACM Transactions on Graphics
  (TOG)}}, Vol.~\bibinfo{volume}{27}. ACM, \bibinfo{pages}{59}.
\newblock


\bibitem[\protect\citeauthoryear{Huang}{Huang}{1997}]%
        {Hua97}
\bibfield{author}{\bibinfo{person}{Zhexue Huang}.}
  \bibinfo{year}{1997}\natexlab{}.
\newblock \showarticletitle{Clustering large data sets with mixed numeric and
  categorical values}. In \bibinfo{booktitle}{{\em Proceedings of the 1st
  pacific-asia conference on knowledge discovery and data mining,(PAKDD)}}.
  Singapore, \bibinfo{pages}{21--34}.
\newblock


\bibitem[\protect\citeauthoryear{Huang}{Huang}{1998}]%
        {Hua98}
\bibfield{author}{\bibinfo{person}{Zhexue Huang}.}
  \bibinfo{year}{1998}\natexlab{}.
\newblock \showarticletitle{Extensions to the k-means algorithm for clustering
  large data sets with categorical values}.
\newblock \bibinfo{journal}{{\em Data mining and knowledge discovery\/}}
  \bibinfo{volume}{2}, \bibinfo{number}{3} (\bibinfo{year}{1998}),
  \bibinfo{pages}{283--304}.
\newblock


\bibitem[\protect\citeauthoryear{Jones, Oliphant, Peterson,
  et~al\mbox{.}}{Jones et~al\mbox{.}}{2001}]%
        {JOP*01}
\bibfield{author}{\bibinfo{person}{Eric Jones}, \bibinfo{person}{Travis
  Oliphant}, \bibinfo{person}{Pearu Peterson}, {et~al\mbox{.}}}
  \bibinfo{year}{2001}\natexlab{}.
\newblock \bibinfo{title}{{SciPy}: Open source scientific tools for {Python}}.
\newblock   (\bibinfo{year}{2001}).
\newblock
\showURL{%
\url{http://www.scipy.org/}}


\bibitem[\protect\citeauthoryear{Karis}{Karis}{2013}]%
        {Kar13}
\bibfield{author}{\bibinfo{person}{Brian Karis}.}
  \bibinfo{year}{2013}\natexlab{}.
\newblock \showarticletitle{Real shading in unreal engine 4}. In
  \bibinfo{booktitle}{{\em ACM SIGGRAPH 2013 Courses: Physically based shading
  in theory and practice}}. \bibinfo{publisher}{ACM}.
\newblock


\bibitem[\protect\citeauthoryear{Kolb, Mitchell, and Hanrahan}{Kolb
  et~al\mbox{.}}{1995}]%
        {KMH95}
\bibfield{author}{\bibinfo{person}{Craig Kolb}, \bibinfo{person}{Don Mitchell},
  {and} \bibinfo{person}{Pat Hanrahan}.} \bibinfo{year}{1995}\natexlab{}.
\newblock \showarticletitle{A realistic camera model for computer graphics}. In
  \bibinfo{booktitle}{{\em SIGGRAPH}}, Vol.~\bibinfo{volume}{95}.
  \bibinfo{pages}{317--324}.
\newblock


\bibitem[\protect\citeauthoryear{Lagarde and de~Rousiers}{Lagarde and
  de~Rousiers}{2014}]%
        {Lag14}
\bibfield{author}{\bibinfo{person}{S{\'e}bastien Lagarde} {and}
  \bibinfo{person}{Charles de Rousiers}.} \bibinfo{year}{2014}\natexlab{}.
\newblock \showarticletitle{Moving frostbite to physically based rendering}. In
  \bibinfo{booktitle}{{\em ACM SIGGRAPH 2014 Courses: Physically based shading
  in theory and practice}}. \bibinfo{publisher}{ACM}.
\newblock


\bibitem[\protect\citeauthoryear{Lensch, Kautz, Goesele, Heidrich, and
  Seidel}{Lensch et~al\mbox{.}}{2003}]%
        {LKG*03}
\bibfield{author}{\bibinfo{person}{Hendrik Lensch}, \bibinfo{person}{Jan
  Kautz}, \bibinfo{person}{Michael Goesele}, \bibinfo{person}{Wolfgang
  Heidrich}, {and} \bibinfo{person}{Hans-Peter Seidel}.}
  \bibinfo{year}{2003}\natexlab{}.
\newblock \showarticletitle{Image-based reconstruction of spatial appearance
  and geometric detail}.
\newblock \bibinfo{journal}{{\em ACM Transactions on Graphics (TOG)\/}}
  \bibinfo{volume}{22}, \bibinfo{number}{2} (\bibinfo{year}{2003}),
  \bibinfo{pages}{234--257}.
\newblock


\bibitem[\protect\citeauthoryear{Li, Foo, Torrance, and Westin}{Li
  et~al\mbox{.}}{2006}]%
        {LFTW06}
\bibfield{author}{\bibinfo{person}{Hongsong Li}, \bibinfo{person}{Sing-Choong
  Foo}, \bibinfo{person}{Kenneth~E Torrance}, {and} \bibinfo{person}{Stephen~H
  Westin}.} \bibinfo{year}{2006}\natexlab{}.
\newblock \showarticletitle{Automated three-axis gonioreflectometer for
  computer graphics applications}.
\newblock \bibinfo{journal}{{\em Optical Engineering\/}} \bibinfo{volume}{45},
  \bibinfo{number}{4} (\bibinfo{year}{2006}), \bibinfo{pages}{043605}.
\newblock


\bibitem[\protect\citeauthoryear{Li, Dong, Peers, and Tong}{Li
  et~al\mbox{.}}{2017}]%
        {LDPT17}
\bibfield{author}{\bibinfo{person}{Xiao Li}, \bibinfo{person}{Yue Dong},
  \bibinfo{person}{Pieter Peers}, {and} \bibinfo{person}{Xin Tong}.}
  \bibinfo{year}{2017}\natexlab{}.
\newblock \showarticletitle{Modeling surface appearance from a single
  photograph using self-augmented convolutional neural networks}.
\newblock \bibinfo{journal}{{\em ACM Transactions on Graphics (TOG)\/}}
  \bibinfo{volume}{36}, \bibinfo{number}{4} (\bibinfo{year}{2017}),
  \bibinfo{pages}{45}.
\newblock


\bibitem[\protect\citeauthoryear{Li, Sunkavalli, and Chandraker}{Li
  et~al\mbox{.}}{2018a}]%
        {LSC18}
\bibfield{author}{\bibinfo{person}{Zhengqin Li}, \bibinfo{person}{Kalyan
  Sunkavalli}, {and} \bibinfo{person}{Manmohan Chandraker}.}
  \bibinfo{year}{2018}\natexlab{a}.
\newblock \showarticletitle{Materials for masses: SVBRDF acquisition with a
  single mobile phone image}. In \bibinfo{booktitle}{{\em Proceedings of the
  European Conference on Computer Vision (ECCV)}}. \bibinfo{pages}{72--87}.
\newblock


\bibitem[\protect\citeauthoryear{Li, Xu, Ramamoorthi, Sunkavalli, and
  Chandraker}{Li et~al\mbox{.}}{2018b}]%
        {LXR*18}
\bibfield{author}{\bibinfo{person}{Zhengqin Li}, \bibinfo{person}{Zexiang Xu},
  \bibinfo{person}{Ravi Ramamoorthi}, \bibinfo{person}{Kalyan Sunkavalli},
  {and} \bibinfo{person}{Manmohan Chandraker}.}
  \bibinfo{year}{2018}\natexlab{b}.
\newblock \showarticletitle{Learning to reconstruct shape and spatially-varying
  reflectance from a single image}. In \bibinfo{booktitle}{{\em SIGGRAPH Asia
  2018 Technical Papers}}. ACM, \bibinfo{pages}{269}.
\newblock


\bibitem[\protect\citeauthoryear{Mukaigawa, Sumino, and Yagi}{Mukaigawa
  et~al\mbox{.}}{2007}]%
        {MSY07}
\bibfield{author}{\bibinfo{person}{Yasuhiro Mukaigawa}, \bibinfo{person}{Kohei
  Sumino}, {and} \bibinfo{person}{Yasushi Yagi}.}
  \bibinfo{year}{2007}\natexlab{}.
\newblock \showarticletitle{Multiplexed illumination for measuring BRDF using
  an ellipsoidal mirror and a projector}. In \bibinfo{booktitle}{{\em Asian
  Conference on Computer Vision}}. Springer, \bibinfo{pages}{246--257}.
\newblock


\bibitem[\protect\citeauthoryear{Stokes, Anderson, Chandrasekar, and
  Motta}{Stokes et~al\mbox{.}}{1996}]%
        {SACM96}
\bibfield{author}{\bibinfo{person}{Michael Stokes}, \bibinfo{person}{Matthew
  Anderson}, \bibinfo{person}{Srinivasan Chandrasekar}, {and}
  \bibinfo{person}{Ricardo Motta}.} \bibinfo{year}{1996}\natexlab{}.
\newblock \showarticletitle{A standard default color space for the
  internet-srgb}.
\newblock \bibinfo{journal}{{\em Microsoft and Hewlett-Packard Joint Report\/}}
  (\bibinfo{year}{1996}).
\newblock


\bibitem[\protect\citeauthoryear{Van Der~Walt, Colbert, and Varoquaux}{Van
  Der~Walt et~al\mbox{.}}{2011}]%
        {vdWCV11}
\bibfield{author}{\bibinfo{person}{Stefan Van Der~Walt},
  \bibinfo{person}{S~Chris Colbert}, {and} \bibinfo{person}{Gael Varoquaux}.}
  \bibinfo{year}{2011}\natexlab{}.
\newblock \showarticletitle{The NumPy array: a structure for efficient
  numerical computation}.
\newblock \bibinfo{journal}{{\em Computing in Science \& Engineering\/}}
  \bibinfo{volume}{13}, \bibinfo{number}{2} (\bibinfo{year}{2011}),
  \bibinfo{pages}{22}.
\newblock


\bibitem[\protect\citeauthoryear{Ward}{Ward}{2003}]%
        {War03}
\bibfield{author}{\bibinfo{person}{Greg Ward}.}
  \bibinfo{year}{2003}\natexlab{}.
\newblock \showarticletitle{Fast, robust image registration for compositing
  high dynamic range photographs from hand-held exposures}.
\newblock \bibinfo{journal}{{\em Journal of graphics tools\/}}
  \bibinfo{volume}{8}, \bibinfo{number}{2} (\bibinfo{year}{2003}),
  \bibinfo{pages}{17--30}.
\newblock


\bibitem[\protect\citeauthoryear{Ward}{Ward}{1992}]%
        {War92}
\bibfield{author}{\bibinfo{person}{Gregory~J Ward}.}
  \bibinfo{year}{1992}\natexlab{}.
\newblock \showarticletitle{Measuring and modeling anisotropic reflection}.
\newblock \bibinfo{journal}{{\em Computer Graphics\/}} \bibinfo{volume}{26},
  \bibinfo{number}{2} (\bibinfo{year}{1992}), \bibinfo{pages}{265--272}.
\newblock


\bibitem[\protect\citeauthoryear{Wold, Esbensen, and Geladi}{Wold
  et~al\mbox{.}}{1987}]%
        {WEG87}
\bibfield{author}{\bibinfo{person}{Svante Wold}, \bibinfo{person}{Kim
  Esbensen}, {and} \bibinfo{person}{Paul Geladi}.}
  \bibinfo{year}{1987}\natexlab{}.
\newblock \showarticletitle{Principal component analysis}.
\newblock \bibinfo{journal}{{\em Chemometrics and intelligent laboratory
  systems\/}} \bibinfo{volume}{2}, \bibinfo{number}{1-3}
  (\bibinfo{year}{1987}), \bibinfo{pages}{37--52}.
\newblock


\bibitem[\protect\citeauthoryear{Zhu, Byrd, Lu, and Nocedal}{Zhu
  et~al\mbox{.}}{1997}]%
        {ZBLN97}
\bibfield{author}{\bibinfo{person}{Ciyou Zhu}, \bibinfo{person}{Richard~H
  Byrd}, \bibinfo{person}{Peihuang Lu}, {and} \bibinfo{person}{Jorge Nocedal}.}
  \bibinfo{year}{1997}\natexlab{}.
\newblock \showarticletitle{Algorithm 778: L-BFGS-B: Fortran subroutines for
  large-scale bound-constrained optimization}.
\newblock \bibinfo{journal}{{\em ACM Transactions on Mathematical Software
  (TOMS)\/}} \bibinfo{volume}{23}, \bibinfo{number}{4} (\bibinfo{year}{1997}),
  \bibinfo{pages}{550--560}.
\newblock


\end{thebibliography}



\appendix
\section{BRDF model}
We elaborate on the Disney BRDF model~\cite{Bur12} in Section~\ref{render}. This simplified implementation ignores parameters modeling subsurface scattering, clearcoat layer and sheen component intended for cloth, yet still covers the majority of common opaque materials. Let $\mathbf{l}$, $\mathbf{v}$, $\mathbf{t}$, $\mathbf{n}$ be direction to light, direction to view, tangent and normal respectively. Half vector $\mathbf{h}=\frac{\mathbf{l}+\mathbf{v}}{||\mathbf{l}+\mathbf{v}||}$ and bitangent $\mathbf{b}=\mathbf{n}\times\mathbf{t}$ are also used. Note these are all normalized vectors. The BRDF $f$ is composed of diffuse and specular part, with details listed below:
\begin{equation}
\begin{aligned}
F_{D90} &= 0.5 + 2(\mathbf{l}\cdot\mathbf{h})^2roughness \\
f_{diff} &= \frac{baseColor(1-metallic)}{\pi} \\
            &~~~~\cdot (1+(F_{D90}-1)(1-\mathbf{l}\cdot\mathbf{n})^5)(1+(F_{D90}-1)(1-\mathbf{v}\cdot\mathbf{n})^5) \\
c_{tint} &= \frac{baseColor}{0.3baseColor_R + 0.6baseColor_G + 0.1baseColor_B} \\
c_{nonmetal} &= 0.08specular(c_{tint}specularTint+(1-specularTint)) \\
c_{spec} &= c_{nonmetal}(1-metallic) + baseColor\cdot metallic \\
\alpha_x &= \max(0.001, roughness^2 / \sqrt{1-0.9anisotropic}) \\
\alpha_y &= \max(0.001, roughness^2 \cdot \sqrt{1-0.9anisotropic}) \\
D_s &= \frac{1}{\pi\alpha_x\alpha_y((\frac{\mathbf{h}\cdot\mathbf{t}}{\alpha_x})^2 + (\frac{\mathbf{h}\cdot\mathbf{b}}{\alpha_y})^2+(\mathbf{h}\cdot\mathbf{n})^2)^2} \\
F_s &= (1-(1-\mathbf{l}\cdot\mathbf{h})^5)c_{spec} + (1-\mathbf{l}\cdot\mathbf{h})^5 \\
G_s &= \frac{1}{\mathbf{l}\cdot\mathbf{n}+\sqrt{(\alpha_x\mathbf{l}\cdot\mathbf{t})^2+(\alpha_y\mathbf{l}\cdot\mathbf{b})^2+(\mathbf{l}\cdot\mathbf{n})^2}} \\
      &~~~~\cdot \frac{1}{\mathbf{v}\cdot\mathbf{n}+\sqrt{(\alpha_x\mathbf{v}\cdot\mathbf{t})^2+(\alpha_y\mathbf{v}\cdot\mathbf{b})^2+(\mathbf{v}\cdot\mathbf{n})^2}} \\
f_{spec} &= D_s F_s G_s \\
f &= f_{diff} + f_{spec}
\end{aligned}
\end{equation}

\end{document}